\documentclass[fleqn,usenatbib]{mnras}

\usepackage[T1]{fontenc}
\usepackage{ae,aecompl}

\usepackage{graphicx}	
\usepackage{amsmath}	
\usepackage{amsfonts}   
\usepackage{amssymb}	
\usepackage{gensymb}    
\usepackage{soul} 
\usepackage{pdflscape}  
\usepackage{booktabs}   
\usepackage{nicefrac}   

\newcommand{\kms} {$\mathrm{ km \; s^{-1}}\,$}

\newcommand{\ergs} {erg s$^{-1}$}
\newcommand{\msol} {M$_{\odot}$}
\newcommand{\zsol} {Z$_{\odot}$}
\newcommand{\rsol} {R$_{\odot}$}

\def\lesssim{\mathrel{\hbox{\rlap{\hbox{\lower4pt\hbox{$\sim$}}}\hbox{$<$}}}}
\def\gtrsim{\mathrel{\hbox{\rlap{\hbox{\lower4pt\hbox{$\sim$}}}\hbox{$>$}}}}
\newcommand{\ang} {\r{A}}
\newcommand{\halpha} {$\mathrm{H\alpha}$}
\newcommand{\hbeta} {$\mathrm{H\beta}\,$}
\newcommand{\hgamma} {$\mathrm{H\gamma}\,$}
\newcommand{\hdelta} {$\mathrm{H\delta}\,$}

\long\def\symbolfootnote[#1]#2{\begingroup%
\def\thefootnote{\fnsymbol{footnote}}\footnote[#1]{#2}\endgroup}

\title[Weighing Melnick 34]{Weighing Melnick 34: the most massive binary system known}

\author[K. Tehrani et al.]{
Katie A. Tehrani,$^{1}$\thanks{k.tehrani@sheffield.ac.uk}
Paul A. Crowther,$^{1}$
Joachim M. Bestenlehner,$^{1}$
\newauthor
Stuart P. Littlefair,$^{1}$
A. M. T. Pollock,$^{1}$
Richard J. Parker,$^{1}$\thanks{Royal Society Dorothy Hodgkin Fellow.} 
\& Olivier Schnurr$^{2}$ \\
$^{1}$Department of Physics and Astronomy, University of Sheffield, Sheffield, S3 7RH, UK \\
$^{2}$Cherenkov Telescope Array Observatory gGmbH, Via Piero Gobetti 93/3, I-40126 Bologna, Italy
}

\date{Accepted XXX. Received YYY; in original form ZZZ}

\pubyear{2018}

\begin{document}
\label{firstpage}
\pagerange{\pageref{firstpage}--\pageref{lastpage}}
\maketitle

\begin{abstract}

Here we confirm Melnick 34, an X-ray bright star in the 30 Doradus region of the Large Magellanic Cloud, as an SB2 binary comprising WN5h+WN5h components. We present orbital solutions using 26 epochs of VLT/UVES spectra and 22 epochs of archival Gemini/GMOS spectra. Radial-velocity monitoring and automated template fitting methods both reveal a similar high eccentricity system with a mass ratio close to unity, and an orbital period in agreement with the 155.1 $\pm$ 1 day X-ray light curve period previously derived by Pollock et al. Our favoured solution derived an eccentricity of 0.68 $\pm$ 0.02 and mass ratio of 0.92 $\pm$ 0.07, giving minimum masses of M$_{\text{A}}$sin$^{3}$(i) = 65 $\pm$ 7~\msol\ and M$_{\text{B}}$sin$^{3}$(i) = 60 $\pm$ 7~\msol. Spectral modelling using WN5h templates with {\sc cmfgen} reveals temperatures of T $\sim$53~kK for each component and luminosities of log(L$_{\text{A}}$/L$_{\odot}$) = 6.43 $\pm$ 0.08 and log(L$_{\text{B}}$/L$_{\odot}$) = 6.37 $\pm$ 0.08, from which BONNSAI evolutionary modelling gives masses of M$_{\text{A}}$ = 139$^{+21}_{-18}$~\msol\ and M$_{\text{B}}$ = 127$^{+17}_{-17}$~\msol\ and ages of $\sim$0.6~Myrs. Spectroscopic and dynamic masses would agree if Mk34 has an inclination of i $\sim$50$\degree$, making Mk34 the most massive binary known and an excellent candidate for investigating the properties of colliding wind binaries. Within 2-3 Myrs, both components of Mk34 are expected to evolve to stellar mass black holes which, assuming the binary system survives, would make Mk34 a potential binary black hole merger progenitor and gravitational wave source.

\end{abstract}




\section{Introduction} 
\label{sec:introduction}

Mass is a fundamental property of stars but is often elusive.
Robust stellar mass determinations are crucial for deducing a wealth of other stellar properties, such as luminosities and lifetimes. However, inferring stellar masses is often a difficult process and direct measurements are rare since reliable model-independent masses can only be obtained from either astrometric binaries, for nearby objects with a known distance, or spectroscopic binary systems in which the orbital inclination is known, i.e. from double-eclipsing binary systems or polarimetry.
The difficulty is amplified when working in the high stellar mass range since massive stars are inherently rare. 
The initial mass function (IMF) describes the distribution of stellar masses and shows that the formation of very massive stars (VMS, $>$100~\msol) is exceptionally rare \citep{bookvin2015}. It has also been claimed that an upper stellar mass limit of $\sim$150~\msol\ exists for the formation of single stars \citep{wei2004, fig2005, oey2005}, however this argument is based on observations rather than physical constraints. Whilst higher mass stars may be possible, it has been suggested that these stars stem from mergers of lower mass stars rather than the formation of VMS \citep{ban2012}.

Due to these complications, we often turn to spectroscopic and evolutionary modelling (e.g. \citealt{yus2013, koh2015}) to determine stellar masses. In the high mass regime however, this can lead to ambiguous and controversial results \citep{cro2010, sch2018}.

To remedy this we need to verify these models with a robust sample of directly measured masses to act as calibration points. The detached eclipsing binaries catalogue, DEBCat \citep{sou2015}, collates a sample of well-studied spectroscopic binary systems with mass and radius measurements accurate to 2\%. Massive stars however, are poorly represented in both visual and spectroscopic searches, with only 3 systems with primary stellar masses greater than 20~\msol\ and none greater than 30~\msol\ appearing in DEBCat.

For VMS, only the double-lined eclipsing binary NGC3603-A1 located in the Galactic cluster NGC 3603 is known to date. This system comprises two WN6ha stars orbiting with a 3.772 day period \citep{mof1984} and radial-velocity monitoring from \citet{sch2008} found direct mass estimates of M$_{\text{A}}$=116 $\pm$ 31~\msol\ and M$_{\text{B}}$=89 $\pm$ 16~\msol\ for the primary and secondary, respectively. 

Within the Large Magellanic Cloud (LMC) a VMS binary has yet to be identified. The most massive double-eclipsing binary currently known is R136-38, an O3~V + O6~V system hosting minimum masses of M$_{\text{A}}$sin$^{3}$(i)=53.8 $\pm$ 0.2~\msol\ and M$_{\text{B}}$sin$^{3}$(i)=22.1 $\pm$ 0.1~\msol\ respectively \citep{mas2002b}. Both spectroscopic and photometric monitoring reveal a 3.39 day period system, with zero eccentricity and high inclination of 79$\degree$ $\pm$ 1$\degree$. 

\begin{figure}
	\includegraphics[width=\columnwidth]{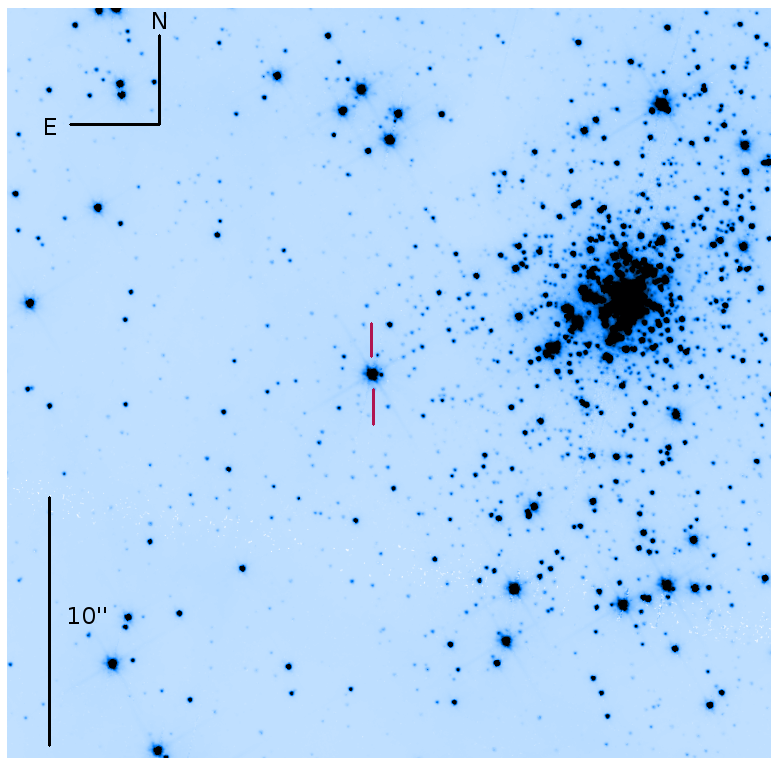}
	\caption{HST/WFC3 F555W image of the 30 Doradus region showing the position of Mk34 relative to the central R136 cluster \citep{dem2014a}. Field of view shown is 30\arcsec $\times$ 30\arcsec, or 7.3~pc $\times$ 7.3~pc, assuming a distance of 50~kpc}
	\label{fig:FoV}
\end{figure}

Melnick 34 (Mk34, BAT99 116, Brey 84), located in the 30 Doradus (30 Dor) region of the LMC, is an X-ray colliding wind binary consisting of a WN5h primary star and an unknown massive companion \citep{tow2006, sch2008, cro2011}. It was revealed as a double-line spectroscopic binary (SB2) by \citet{che2011} who obtained a partial radial-velocity curve during brief spectroscopic monitoring over 51 days starting in late 2009.

Mk34 is located on the periphery of R136, the rich star cluster within 30 Dor, at a projected distance of $\sim$2~pc from the central star R136a1 as shown in Fig.~\ref{fig:FoV}, and is one of the most luminous stars in the LMC, with a luminosity of log(L/L$_{\odot}$)=7.05 and an inferred mass of M=390~\msol\ \citep{hai2014} using the theoretical M/L relationship from \citet{gra2011}.

X-ray observations revealed a more impressive picture, showcasing Mk34 as the most luminous X-ray source in the nebula, with log(L$_{\text{x}}$)=35.2~\ergs\ \citep{tow2006}. The high X-ray/bolometric luminosity ratio, L$_{\text{x}}$/L$_{\text{bol}}$ = 3.7 $\times$ 10$^{-6}$, prompted the suspicion that Mk34 is a colliding wind binary system. Recurrent X-ray variability, however, had not been observed until recently. The Chandra `X-ray Visionary Programme' the Tarantula -- Revealed by X-rays (T-ReX, P.I. Townsley) provided extensive X-ray coverage of the region over $\sim$700 days. This long baseline, combined with frequent observations, was sufficient to reveal the 155.1 $\pm$ 1 day X-ray cycle of Mk34 and further support the binary hypothesis \citep{pol2018}.

To settle this issue and investigate the system further requires optical data: confirmation of the binary status of the system would provide a unique opportunity to obtain direct mass estimates for VMS in a low-metallicity environment.

Here we report new VLT/UVES optical spectroscopic monitoring of Melnick 34 across a full orbital period, supplemented by archival Gemini/GMOS spectra \citep{che2011}. In Sect.~\ref{sec:orb_sol} we show radial-velocity measurements including the derived orbital properties of the system. In Sect.~\ref{sec:phys_and_wind} we present spectral and evolutionary modelling parameters for the system. In Sect.~\ref{sec:discussion} we discuss the limitations and implications of these results, and give brief conclusions in Sect.~\ref{sec:conclusion}


\section{Observations}
\label{sec:observations}

\begin{figure*}
	\includegraphics[width=\textwidth]{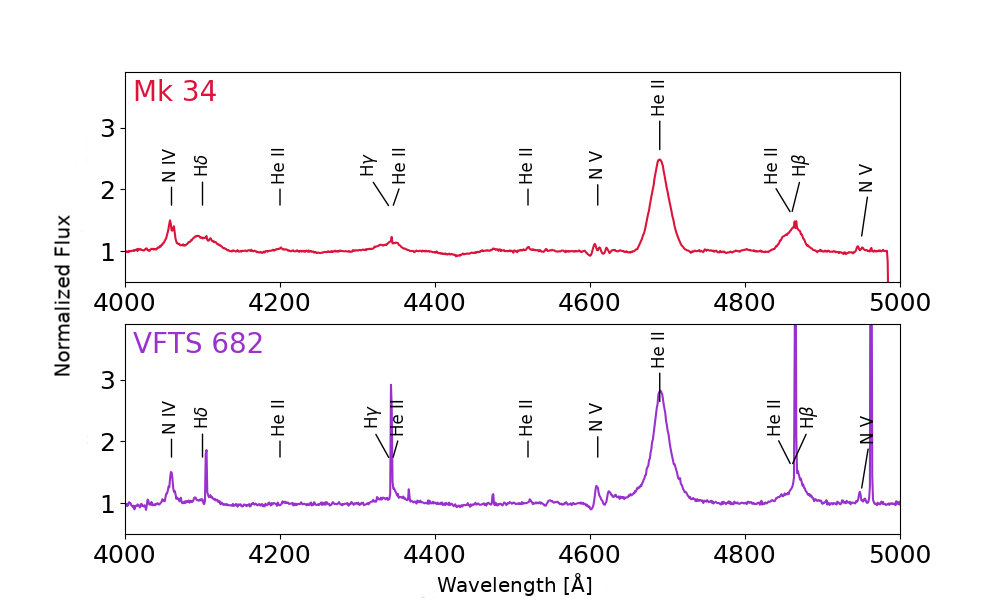}
	\caption{Top panel: Epoch 2 of the VLT/UVES data showing the blue spectrum of Mk34. Double-lined emission peaks can be seen in the N\,{\sc iv} 4058, N\,{\sc v} 4603--20 P Cygni profiles and N\,{\sc v} 4945 emission lines confirming the system is an SB2 binary. Bottom panel: VFTS 682 spectrum from \citet{bes2011, bes2014} classified as a single WN5h star and shown here to demonstrate the similarities between the two spectra, although narrow nebular emission lines are also present here. Relevant emission lines used for classification and radial-velocity measurements have been identified.}
	\label{fig:ob2}
\end{figure*}

\subsection{UVES Observations}
\label{sec:uves}

26 epochs of VLT/UVES spectra were obtained in service mode over a 497 day period -- from 2016 September 6 to 2018 January 16 (Program--ID 098.D-0108(A), PI Crowther). Observations were taken in the Dichroic 2 instrument mode, giving spectral ranges of 373--499~nm (blue arm), 567--757~nm (red arm lower) and 767--945~nm (red arm upper). Exposure time for each observation was 450~s, with the exception of the first Epoch 0 which was a test run with an exposure time of 120~s, and Epoch 3 which had slightly shorter exposure time of 400~s. A 0.8\arcsec\ slit was used throughout. The reciprocal dispersion scale was 0.02--0.03~\ang/pixel and the resolution was $\sim$0.06~\ang\ at 4000~\ang\ and $\sim$0.09~\ang\ at 5000~\ang, with a S/N ratio of 30--40 per resolution element across all epochs. The resultant resolving power was $\sim$60000 ($\sim$5~\kms).

In order to maximise the chance of observing both components of Mk34 whilst still monitoring the system across the full period, the observations were tailored to achieve a high cadence near X-ray maximum, presumed close to periastron, with more sparse sampling during the X-ray quiescent periods. A full list of observations can be found in Table~\ref{tab:OBs_uves_gmos}.

Bias, flat-field and ThAr arc calibrations were obtained through the standard calibration routine for service mode by ESO and the data reduction was performed using the ESO pipeline Reflex \citep{fre2013}. 

\begin{table*}
	\caption{The upper table shows VLT/UVES barycentric radial velocities measured for star A and star B using the N\,{\sc iv} 4058 and N\,{\sc v} 4945 emission lines. Similarly, the lower table shows archival Gemini S/GMOS barycentric radial velocities measured for star A and star B using the same emission lines. All phase measurements were calculated based on the orbital solution UG1, shown in Table~\ref{tab:orbital_parameters}}
	\label{tab:OBs_uves_gmos}
	\begin{tabular}{c c c c c c c}
		\hline\hline
        \multicolumn{7}{c}{\textit{VLT/UVES}} \\
        \hline
		      &          & \multicolumn{2}{c}{N\,{\sc iv} 4058} & \multicolumn{2}{c}{N\,{\sc v} 4945} &  \\
		Epoch & MJD      & RV$ _{\text{A}} $    & RV$ _{\text{B}} $           & RV$ _{\text{A}} $    & RV$ _{\text{B}} $          & Phase \\
		      &          &  \multicolumn{2}{c}{[\kms]}  & \multicolumn{2}{c}{[\kms]}  &  \\ \hline
		  0   & 57637.41  & $263 \pm 75$ & $258 \pm 70$        & $268 \pm 80$ & $283 \pm 75$      & 0.7816  \\
		  1   & 57665.38  & $428 \pm 20$ & $98 \pm 20$         & $488 \pm 40$ & $128 \pm 20$      & 0.9626  \\
		  2   & 57666.25  & $428 \pm 20$ & $88 \pm 20$         & $508 \pm 40$ & $118 \pm 20$      & 0.9682  \\
		  3   & 57667.21  & $468 \pm 20$ & $73 \pm 30$         & $463 \pm 35$ & $78 \pm 30$       & 0.9745  \\
		  4   & 57681.18  & $343 \pm 15$ & $223 \pm 15$        & $318 \pm 50$ & $338 \pm 30$      & 0.0649  \\
		  5   & 57684.25  & $318 \pm 20$ & $198 \pm 20$        & $332 \pm 65$ & $313 \pm 45$      & 0.0848  \\
		  6   & 57687.32  & $253 \pm 65$ & $238 \pm 50$        & $298 \pm 50$ & $313 \pm 35$      & 0.1046  \\
		  7   & 57690.17  & $158 \pm 20$ & $323 \pm 15$        & $328 \pm 60$ & $343 \pm 35$      & 0.1230  \\
		  8   & 57692.22  & $193 \pm 15$ & $328 \pm 10$        & $308 \pm 90$ & $338 \pm 60$      & 0.1363  \\
		  9   & 57694.22  & $178 \pm 20$ & $333 \pm 25$        & $343 \pm 65$ & $343 \pm 65$      & 0.1492  \\
		 10   & 57723.13  & $238 \pm 60$ & $253 \pm 45$        & $293 \pm 35$ & $298 \pm 30$      & 0.3363  \\
		 11   & 57725.12  & $234 \pm 45$ & $239 \pm 40$        & $304 \pm 15$ & $304 \pm 15$      & 0.3491  \\
		 12   & 57744.21  & $244 \pm 55$ & $249 \pm 40$        & $269 \pm 40$ & $289 \pm 20$      & 0.4727  \\
		 13   & 57751.12  & $290 \pm 50$ & $275 \pm 35$        & $330 \pm 50$ & $320 \pm 40$      & 0.5174  \\
		 14   & 57761.29  & $265 \pm 35$ & $255 \pm 25$        & $275 \pm 15$ & $275 \pm 15$      & 0.5832  \\
		 15   & 57797.03  & $266 \pm 75$ & $251 \pm 50$        & $291 \pm 60$ & $281 \pm 50$      & 0.8145  \\
		 16   & 57800.23  & $276 \pm 35$ & $256 \pm 35$        & $311 \pm 40$ & $271 \pm 30$      & 0.8352  \\
		 17   & 57806.14  & $271 \pm 60$ & $266 \pm 35$        & $441 \pm 20$ & $246 \pm 25$      & 0.8734  \\
		 18   & 57808.03  & $331 \pm 30$ & $221 \pm 20$        & $406 \pm 45$ & $251 \pm 30$      & 0.8856  \\
		 19   & 57811.03  & $366 \pm 15$ & $186 \pm 25$        & $386 \pm 45$ & $186 \pm 25$      & 0.9051  \\
		 20   & 57815.13  & $391 \pm 20$ & $101 \pm 20$        & $411 \pm 50$ & $161 \pm 30$      & 0.9316  \\
		 21   & 57817.01  & $406 \pm 25$ & $131 \pm 40$        & $476 \pm 45$ & $161 \pm 30$      & 0.9437  \\
		 22   & 57823.02  & $461 \pm 20$ & $71 \pm 20$         & $501 \pm 20$ & $111 \pm 20$      & 0.9826  \\
		 23   & 57831.01  & $254 \pm 45$ & $244 \pm 35$        & $279 \pm 50$ & $269 \pm 40$      & 0.0344  \\ 
		 24   & 58133.17  & $499 \pm 20$ & $74 \pm 15$         & $509 \pm 10$ & $99 \pm 10$       & 0.9895  \\ 
		 25   & 58134.17  & $519 \pm 10$ & $59 \pm 20$         & $509 \pm 30$ & $69 \pm 10$       & 0.9959  \\ 
         \hline\hline
         \multicolumn{7}{c}{\textit{Gemini S/GMOS}} \\
         \hline
              &          & \multicolumn{2}{c}{N\,{\sc iv} 4058} & \multicolumn{2}{c}{N\,{\sc v} 4945} &       \\
		Epoch & MJD      & RV$ _{\text{A}} $    & RV$ _{\text{B}} $           & RV$ _{\text{A}} $    & RV$ _{\text{B}} $          & Phase \\
		      &          &  \multicolumn{2}{c}{[\kms]}  & \multicolumn{2}{c}{[\kms]}  &       \\ 
		\hline
		  1   & 55189.04  & $399 \pm 40$  & $119 \pm 30$    & $429 \pm 30$  & $169 \pm 20$   & 0.9396  \\
		  2   & 55189.04  & $399 \pm 40$  & $119 \pm 40$    & $429 \pm 60$  & $169 \pm 30$   & 0.9396  \\
		  3   & 55191.21  & $419 \pm 50$  & $99 \pm 90$     & $419 \pm 30$  & $149 \pm 30$   & 0.9536  \\
		  4   & 55195.22  & $489 \pm 100$ & $49 \pm 130$    & $439 \pm 70$  & $69 \pm 30$    & 0.9796  \\
		  5   & 55196.16  & $489 \pm 40$  & $79 \pm 50$     & $459 \pm 60$  & $69 \pm 40$    & 0.9856  \\
		  6   & 55197.32  & $509 \pm 70$  & $-21 \pm 40$    & $459 \pm 60$  & $69 \pm 60$    & 0.9932  \\
		  7   & 55198.29  & $489 \pm 80$  & $49 \pm 50$     & $509 \pm 90$  & $-11 \pm 60$   & 0.9994  \\
		  8   & 55199.21  & $489 \pm 140$ & $59 \pm 130$    & $539 \pm 80$  & $39 \pm 70$    & 0.0054  \\
		  9   & 55207.21  & $270 \pm 100$ & $270 \pm 100$   & $300 \pm 40$  & $300 \pm 40$   & 0.0572  \\
		 10   & 55210.11  & $250 \pm 100$ & $250 \pm 100$   & $320 \pm 60$  & $320 \pm 60$   & 0.0759  \\
		 11   & 55212.03  & $260 \pm 100$ & $260 \pm 100$   & $320 \pm 50$  & $320 \pm 50$   & 0.0883  \\
		 12   & 55213.03  & $270 \pm 110$ & $270 \pm 110$   & $310 \pm 70$  & $310 \pm 70$   & 0.0948  \\
		 13   & 55214.24  & $280 \pm 100$ & $280 \pm 100$   & $310 \pm 70$  & $310 \pm 70$   & 0.1027  \\
		 14   & 55216.16  & $280 \pm 110$ & $280 \pm 110$   & $290 \pm 60$  & $290 \pm 60$   & 0.1150  \\
		 15   & 55225.12  & $250 \pm 110$ & $250 \pm 110$   & $310 \pm 50$  & $310 \pm 50$   & 0.1730  \\
		 16   & 55227.11  & $270 \pm 120$ & $270 \pm 120$   & $310 \pm 60$  & $310 \pm 60$   & 0.1859  \\
		 17   & 55229.21  & $280 \pm 100$ & $280 \pm 100$   & $340 \pm 110$ & $340 \pm 110$  & 0.1995  \\
		 18   & 55236.10  & $221 \pm 200$ & $221 \pm 200$   & $311 \pm 50$  & $311 \pm 50$   & 0.2440  \\
		 19   & 55237.10  & $261 \pm 120$ & $261 \pm 120$   & $301 \pm 60$  & $301 \pm 60$   & 0.2505  \\
		 20   & 55238.12  & $281 \pm 80$  & $281 \pm 80$    & $311 \pm 60$  & $311 \pm 60$   & 0.2571  \\
		 21   & 55241.07  & $251 \pm 200$ & $251 \pm 200$   & $311 \pm 80$  & $311 \pm 80$   & 0.2763  \\
		 22   & 55242.12  & $281 \pm 150$ & $281 \pm 150$   & $291 \pm 70$  & $291 \pm 70$   & 0.2830  \\
		 \hline\hline
	\end{tabular}
\end{table*}

\subsection{GMOS Observations}
\label{sec:gmos}

Archival Gemini-S/GMOS data (Program--ID GS-2009B-Q-32, PI Chen{\'e}) consist of 22 spectra taken between 2009 December 24 and 2010 February 15. Each observation had an exposure time of 400~s and used a fixed slit width of 0.5\arcsec. Here the reciprocal dispersion scale was 0.47~\ang/pixel, and the resolution was found to be $\sim$1.29~\ang\ at 4000~\ang\ and $\sim$1.16~\ang\ at 5000~\ang, corresponding to a much lower resolving power for this data set of $\sim$3~500. The S/N per resolution element was between 110--170 at 5000~\ang, however dropped considerably to 5--30 at 4000~\ang.

The reduction of the GMOS data was completed using the apextract package in {\sc iraf} \citep{tod1986}. A complete list of Gemini/GMOS observations can be found in Table~\ref{tab:OBs_uves_gmos}.


\subsection{Classification}
\label{sec:classification}

\begin{figure*}
	\includegraphics[width=0.95\textwidth]{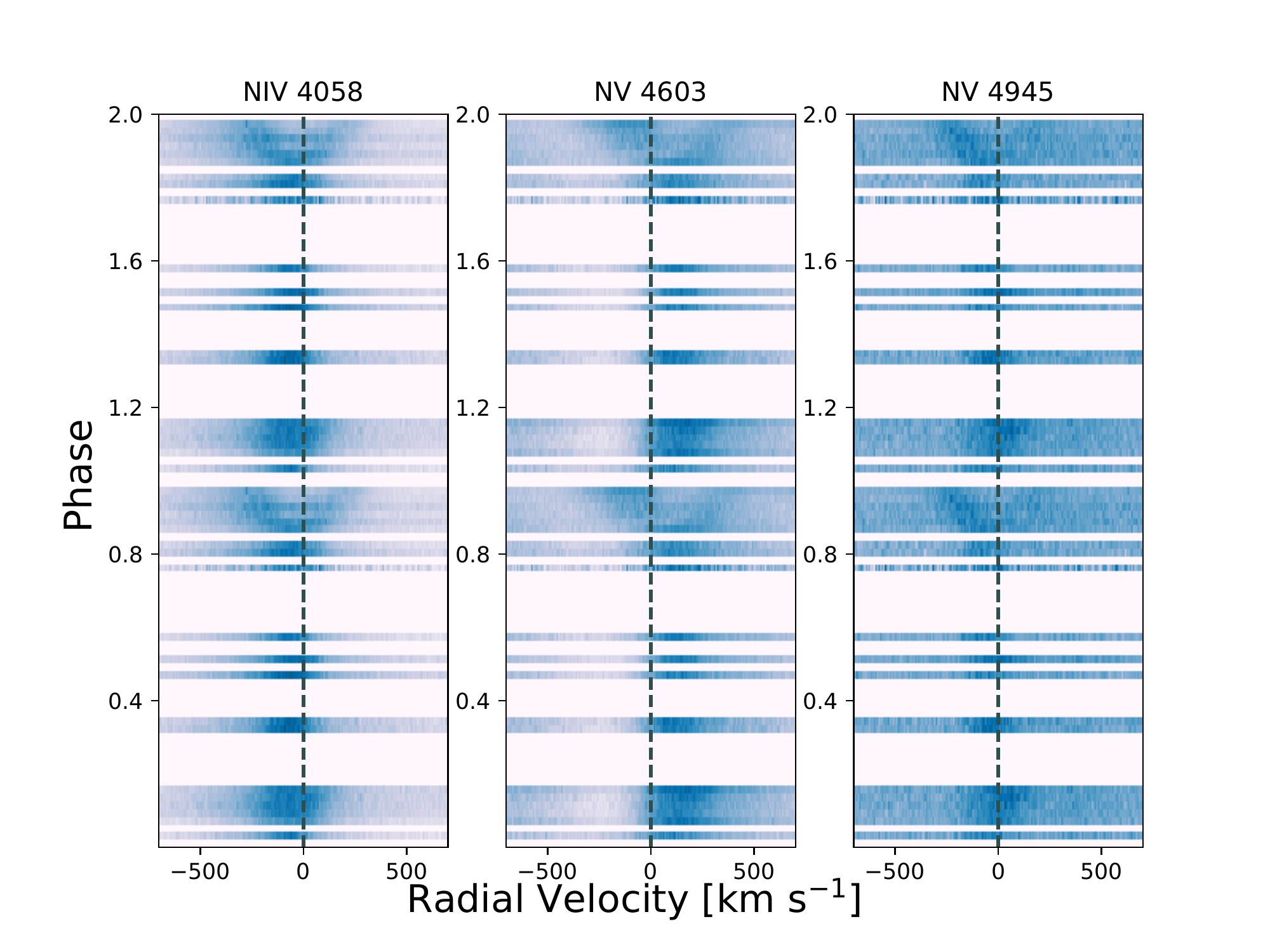}
	\caption{Dynamic representation of the evolution of the N\,{\sc iv} 4057.7, N\,{\sc v} 4603.7 and N\,{\sc v} 4944.6 emission-line radial-velocity profiles from the UVES dataset, across two orbital phases, showing the double emission from the two components of Mk34 at phase 0.8-1, from solution UG1. Radial velocities have been corrected for a systemic velocity of 287~\kms found from solution UG1.}
	\label{fig:trail_plot}
\end{figure*}

Fig.~\ref{fig:ob2} shows a subset of a typical blue UVES spectrum of Mk34 together with a spectrum of the typical WN5h star VFTS 682 from \citet{bes2011, bes2014} for comparison. Both stars show strong, broad He\,{\sc ii} 4686 and Balmer series emission, and sharper N\,{\sc iv} 4058, N\,{\sc v} 4603--20 and N\,{\sc v} 4945 emission lines. Upon closer inspection of these narrow lines, we observe two stellar components in a subset of epochs, as shown in Fig.~\ref{fig:trail_plot}, confirming the SB2 nature of the system. This double emission is brief, occurring in 8 of the 26 epochs observed, covering more than 8 and less than 20 days of the 155 day orbit. For the remainder of the orbit it is not possible to distinguish the two components, suggesting the system is highly eccentric.

The UVES spectra of Mk34 extend beyond this wavelength range, where we also note strong \halpha\ and N\,{\sc iv} 7103--7129 emission, and narrow C\,{\sc iv} 5801--12 emission. Note that the double emission observed at N\,{\sc iv} 7103--7129 can be observed throughout the orbit as this emission is composed of multiple transitions, therefore using this line to trace the stellar radial-velocity curve is difficult and we do not attempt to use it. 

\begin{figure*}
	\includegraphics[width=\textwidth]{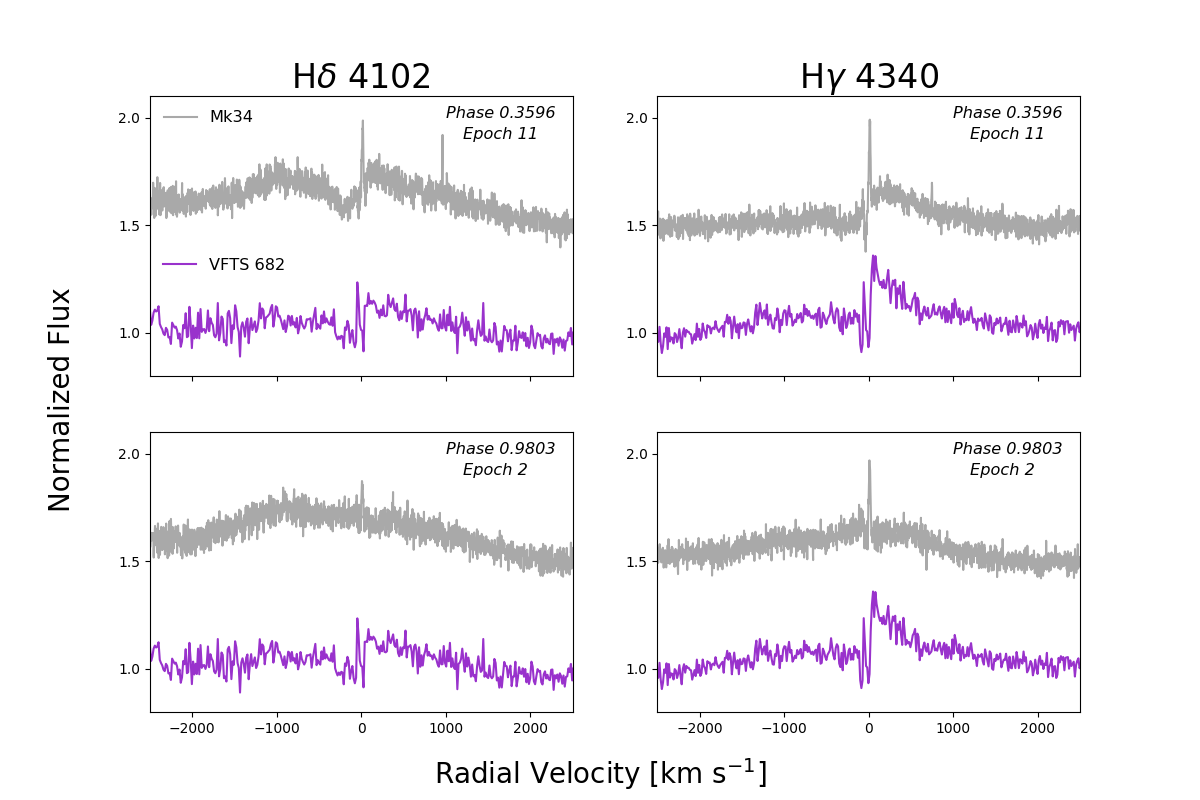}
	\caption{A comparison of the \hdelta\ 4101.7 and \hgamma\ 4340.5 emission line profiles at phases close to periastron (Epoch 2) and apastron (Epoch 11). For comparison, matching emission line profiles from the VFTS 682 spectra have also been plotted \citep{bes2011, bes2014} to highlight the morphological change in the line profile shape visible at periastron. Nebular emission lines have been removed, using Gaussian profile fits, and a 0.5 vertical offset has been applied to the Mk34 spectrum for clarity. The radial-velocities have been corrected for a systemic velocity of 287~\kms found from solution UG1.}
	\label{fig:hdelta_hgamma}
\end{figure*}

When investigating the spectral type of each component we note the N\,{\sc iv} 4058 emission is stronger than the  N\,{\sc v} 4603--20 in both stars, and N\,{\sc iii} 4640 emission is absent, consistent with a WN5 ionization class as outlined in \citet{smi1996, cro2011}. Also, using the Pickering decrement to search for oscillations in the strength of the He\,{\sc ii} (n--4) lines, we note the strong He\,{\sc ii} 4859 + \hbeta\ and He\,{\sc ii} 4340 + \hgamma\ emission, coupled with the weaker He\,{\sc ii} 4541 emission. This indicates a clear hydrogen excess and is characteristic of a WN5h designation \citep{smi1996}. The similarity of the N\,{\sc v}/N\,{\sc iv} ratio for each star suggests the temperatures of these objects are comparable, and these emission features are common to both components, suggesting that both stars are likely to be of the WN5h spectral type.

We also note a phase-dependent change in the morphology of the \hdelta\ and \hgamma\ emission, which is not present in the VFTS 682 spectrum and therefore not intrinsic to the WN5h spectral type. As shown in Fig.~\ref{fig:hdelta_hgamma}, at times close to periastron the emission is smooth. However, near apastron an additional absorption feature develops superimposed upon the emission. We suggest this strong \hdelta\ emission is characteristic of the binary since it is also noted in the known binary R136c \citep{bes2014}, whilst other known single WN5h stars, such as R136a1--3 in the central cluster of 30 Doradus, do not replicate this morphology \citep{cro2010}. In general though, the spectrum of Mk34 is morphologically similar to these other known WN5h stars.


\section{Orbital Solution}
\label{sec:orb_sol}

\subsection{Radial-Velocity Measurement}
\label{sec:rv}

\begin{figure*} 
	\includegraphics[width=\textwidth]{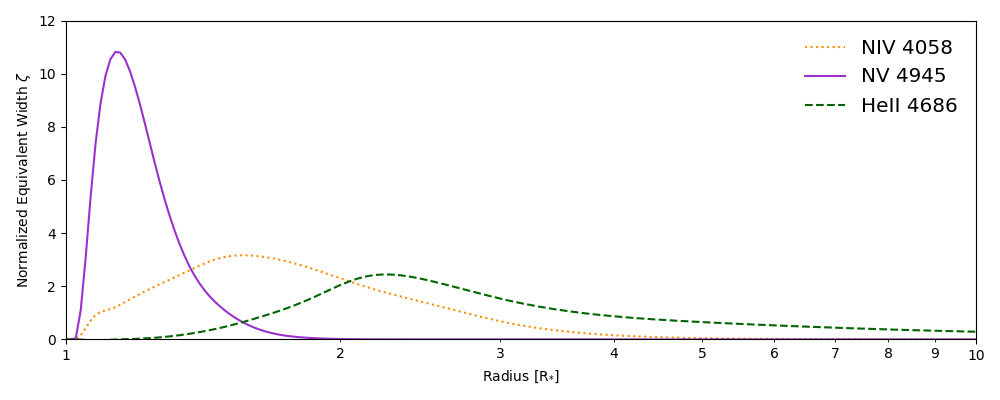}
	\caption{A comparison of the position of the line-forming region for various ions showing that the N\,{\sc iv} 4058 and N\,{\sc v} 4945 emission lines are better tracers of the stellar kinematics than the He\,{\sc ii} 4686 line which is formed further out in the stellar wind.}
	\label{fig:lfr}
\end{figure*}

To reveal the radial-velocity distribution over time we use a template theoretical spectrum of VFTS 682 (WN5h) from \citet{bes2011, bes2014} to measure the N\,{\sc iv} 4058 and N\,{\sc v} 4945 emission-line radial velocities. We are restricted to these lines, which are formed in the wind, because no photospheric lines are available for WR stars. We assume that the centroids of the emission lines in the template models accurately reflect the stellar kinematics. Amongst all potential diagnostics, N\,{\sc iv} 4058 and N\,{\sc v} 4945 originate deep in the stellar wind with line morphologies lacking complications of P Cygni profiles (e.g. N\,{\sc v} 4603-20), therefore we select these lines for our radial-velocity analysis. Fig.~\ref{fig:lfr} shows a comparison of the line-forming regions for various ions, in particular emphasizing the location of the N\,{\sc iv} 4058 and N\,{\sc v} 4945 emission line formation, which is much closer to the stellar photosphere ($\leq$ 1.4~R$_{*}$) than the He\,{\sc ii} 4686 line forming region. 


The template spectrum was used to measure radial-velocities using a simple shift and add method, with each fit being visually examined, for both the UVES and GMOS data sets. The results are recorded in Table~\ref{tab:OBs_uves_gmos}. It is important to note that due to the poor quality of the GMOS data there is a large degree of uncertainty in the radial-velocities measured from this data set, however the longer baseline is useful for helping to constrain the orbital period. We also note a systematic offset between the N\,{\sc iv} 4058 and N\,{\sc v} 4945 radial velocities, with an average offset of 42 $\pm$ 44~\kms in the UVES data and 33 $\pm$ 36~\kms in the GMOS data which we discuss further in Sect.~\ref{sec:spec}.

\begin{figure*}
	\includegraphics[width=\textwidth]{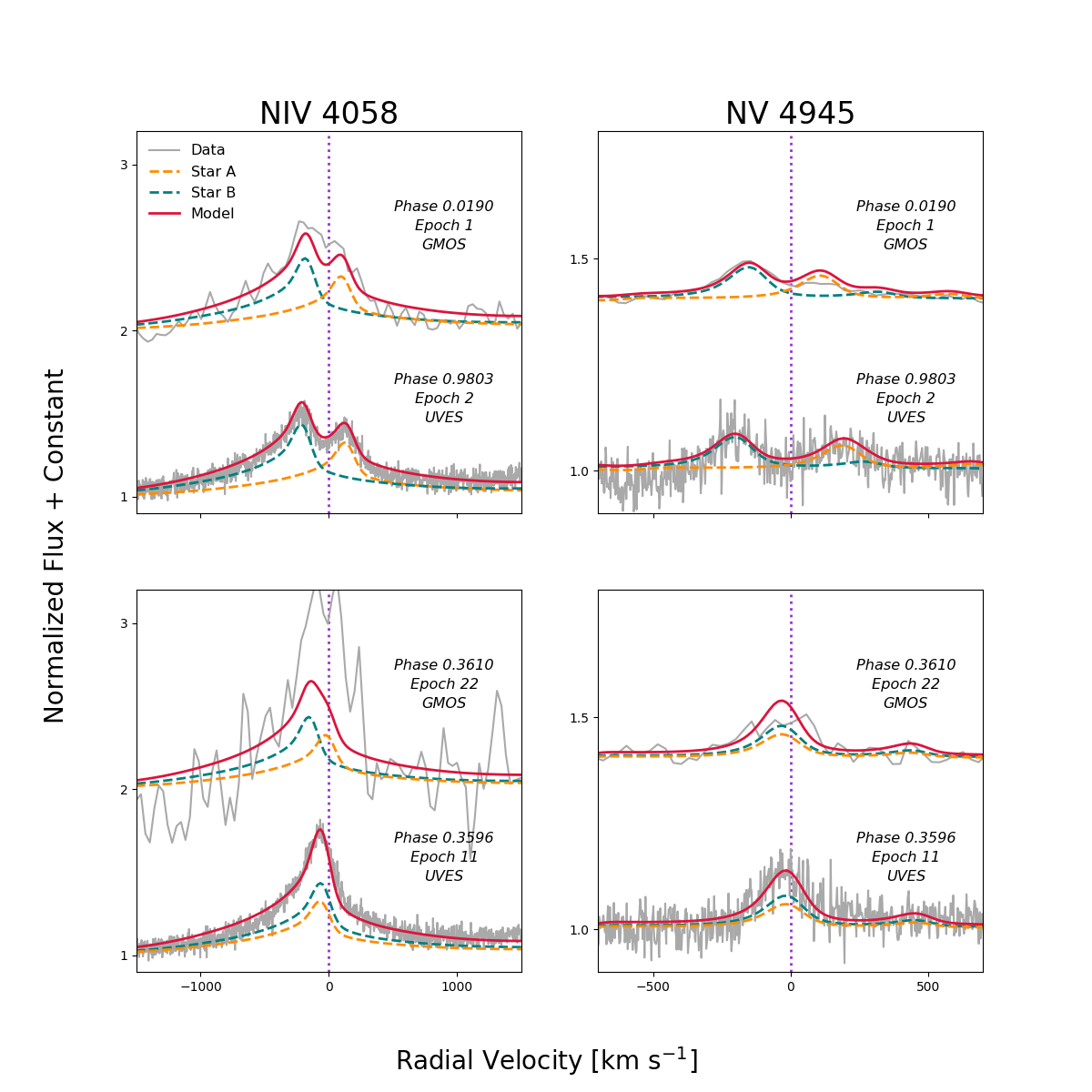}
	\caption{A comparison of the GMOS and UVES radial-velocity modelling for the N\,{\sc iv} 4058 and N\,{\sc v} 4945 emission lines from two different orbital phases to show both double emission and blended epochs. A vertical offset has been introduced to the GMOS data for clarity. The template for both star A and star B has been broadened to account for rotational velocity of $\sim$125~\kms. The applied systemic velocity is 287~\kms found from solution UG1.}
	\label{fig:rv_modelling}
\end{figure*}

Fig.~\ref{fig:rv_modelling} shows an example of this modelling at two extreme stages; at phase $\sim$0 near periastron when the double-lined emission is very clear and at phase $\sim$0.4 when emission lines from both stars are blended.
Due to the significant intrinsic line widths of both stars involved -- FWHM $\sim$300~\kms (N\,{\sc iv} 4058) and FWHM $\sim$250~\kms (N\,{\sc v} 4945) -- measuring the radial velocities away from periastron proved difficult and therefore they have significant uncertainties, typically $\pm$40~\kms for UVES, and $\pm$100~\kms for GMOS.

\subsection{Orbital Properties}
\label{sec:orbital_prop}

\citet{pol2018} established a 155.1 $\pm$ 1 day X-ray cycle which was strongly suspected to be the orbital period. Here we can use a combination of the UVES and GMOS data providing a baseline of over 8 years, translating to 19 orbits, better to constrain this period.

\begin{figure}
	\includegraphics[width=\columnwidth]{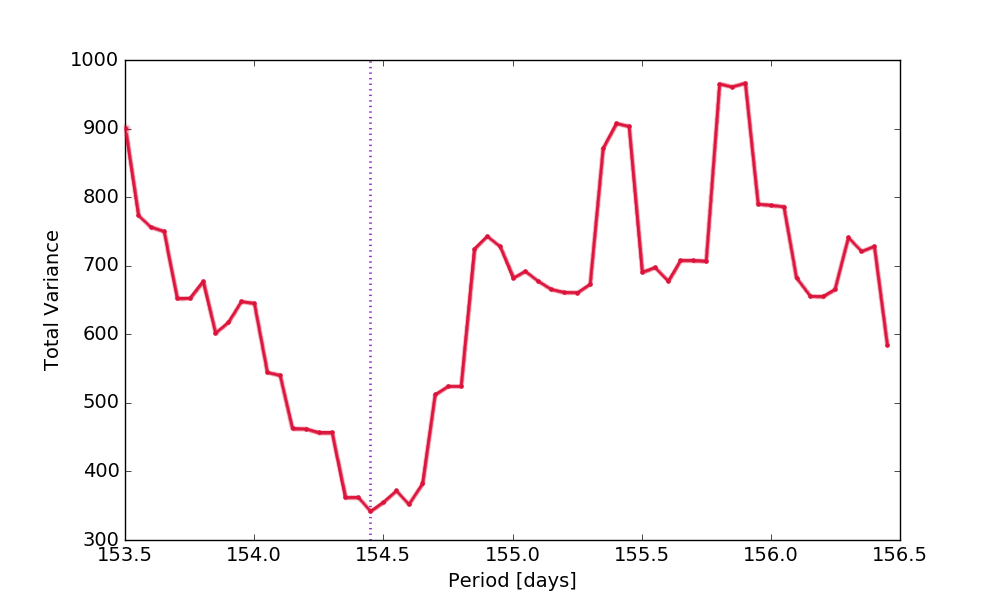}
	\caption{Phase dispersed minimisation plot produced from a combination of VLT/UVES and Gemini/GMOS data showing the total variance minimum corresponds to a period of 154.45 days.}
	\label{fig:pdm}
\end{figure}

To start, we adopt the technique of phase dispersion minimisation (PDM) to identify a range of periods to use as prior probabilities in future analysis. PDM works by varying the period and folding the radial-velocity data in phase before splitting the orbit into bins. We then calculate the variance of the radial velocities within each bin and sum across the full orbit to obtain a total variance. Finally, plotting period against total variance as shown in Fig.~\ref{fig:pdm} will reveal the most likely period. In this instance, using both the UVES and GMOS data with 20 phase bins, the most likely period was 154.45 days.

\begin{figure*}
	\includegraphics[width=0.7\textwidth]{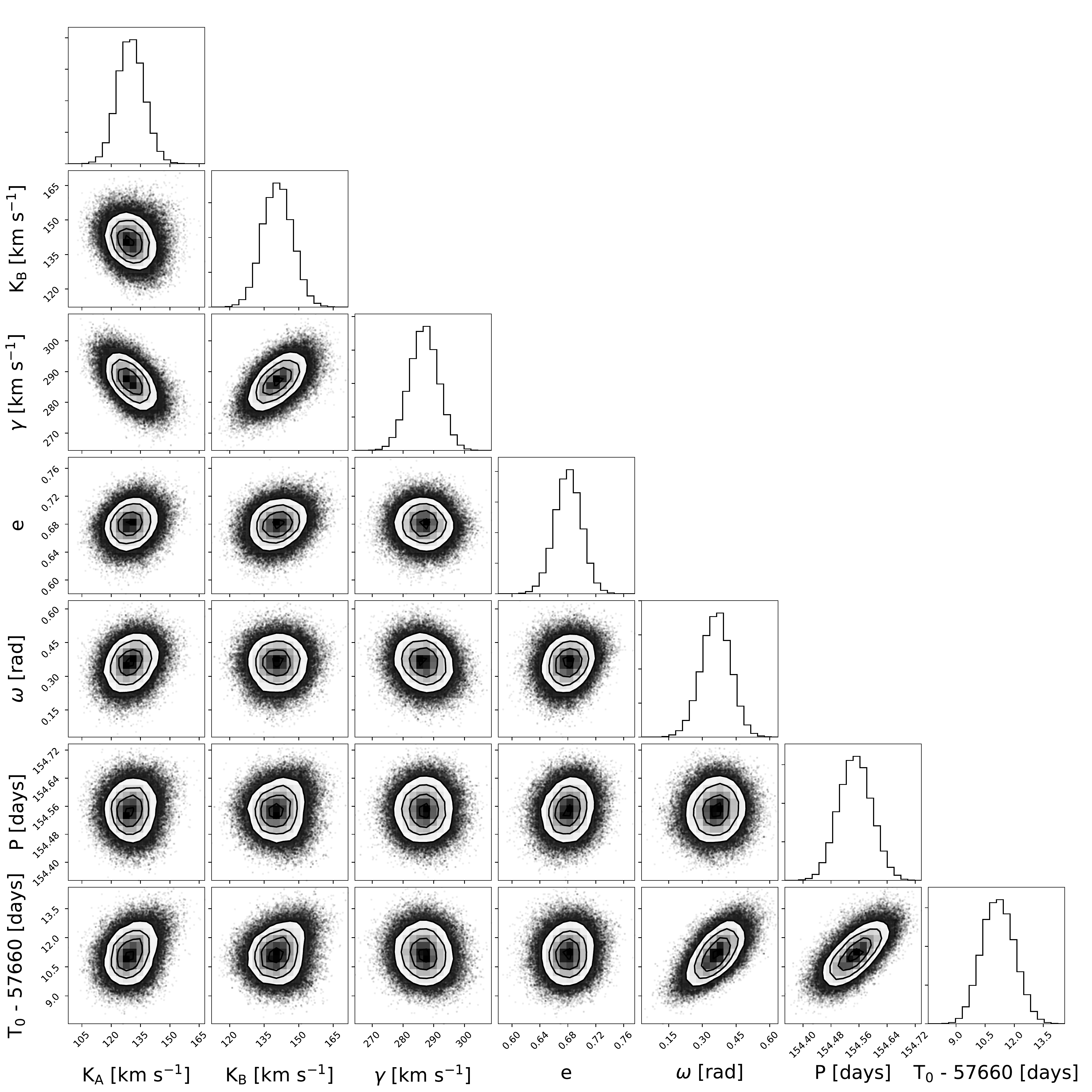}
	\caption{Corner plot showing posterior probabilities for solution UG1, where $K_{\text{A}}$ and $K_{\text{B}}$ are the semi-amplitudes of the velocities for star A and star B respectively, $\gamma$ is the systemic velocity, e is the eccentricity, omega is the longitude of the periastron, P is the orbital period and T$_{0}$ is the time of periastron.}
	\label{fig:cornerplot_gmos_uves}
\end{figure*}

\begin{figure*}
	\includegraphics[width=\textwidth]{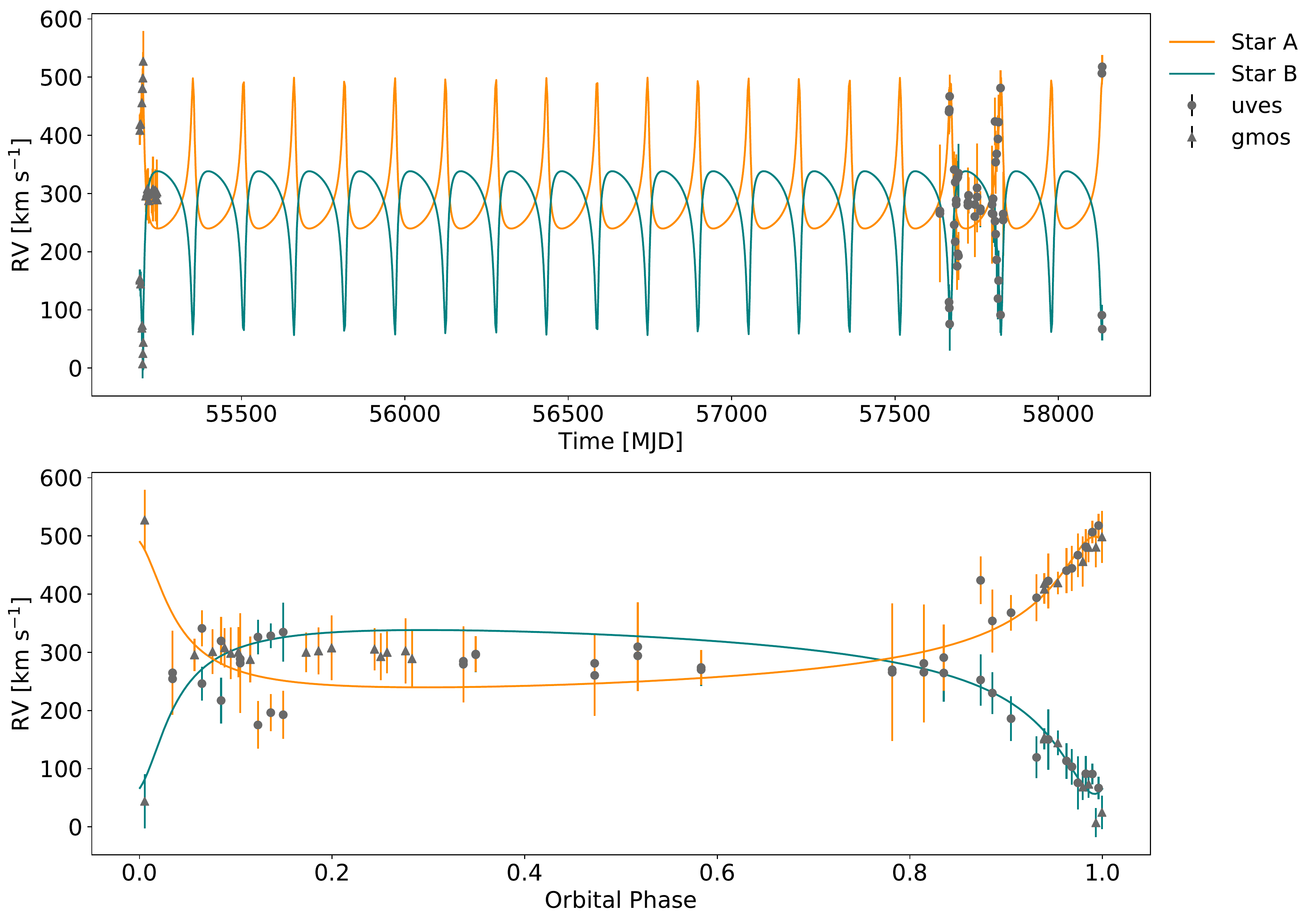}
	\caption{Best fitting radial-velocity curve for the VLT/UVES and Gemini/GMOS data providing the parameters for solution UG1.}
	\label{fig:rvfit_gmos_uves}
\end{figure*}


To obtain a full set of orbital parameters we used Markov Chain Monte Carlo (MCMC) fitting, with a uniform prior distribution for the period of 155 $\pm$ 3 days, guided by the PDM results. During this MCMC fitting we used the averaged radial-velocity measurements of the N\,{\sc iv} 4058 and N\,{\sc v} 4945 emission lines from the UVES data, and the averaged N\,{\sc iv} 4058 and N\,{\sc v} 4945 radial-velocities from the GMOS data, and applied a barycentric correction to provide 48 epochs of observations available for fitting.
The MCMC fitting was carried out twice, once using the combined GMOS and UVES data, and again using only the more reliable UVES measurements, to provide two orbital solutions, solution UG1 and solution U1 respectively. A full set of Keplerian orbital parameters was used in each fit: the semi-amplitudes of the velocities for star A and star B, $K_{\text{A}}$ and $K_{\text{B}}$ respectively; systemic velocity, $\gamma$; eccentricity, e; longitude of periastron, $\omega$; orbital period, P; and time of periastron, T$_{0}$); plus a nuisance parameter which was a multiplier for the formal error bars. Uninformative priors were adopted for all parameters, except the period, which for solution UG1 we adopted a uniform probability distribution prior of 155 $\pm$ 3 days and, based on the subsequent results of this fit, for solution U1 we use a Gaussian prior distribution of 154.5 $\pm$ 0.5 days.

\begin{figure}
	\includegraphics[width=\columnwidth]{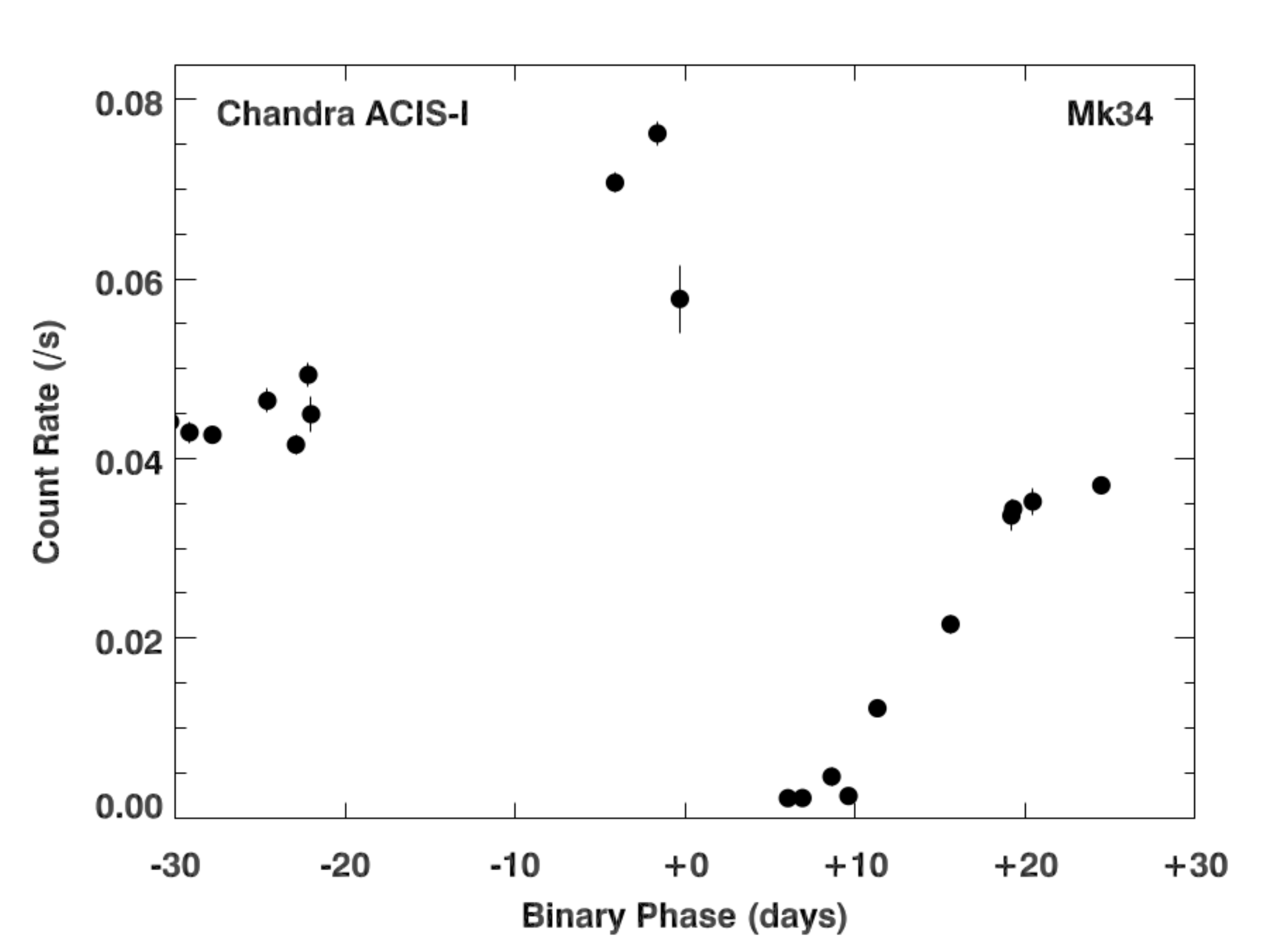}
	\caption{Detailed plot of the X-Ray variability observed in \citet{pol2018} close to periastron, based on the 154.5 day period derived in this work}
	\label{fig:xray_peri}
\end{figure}

Fig.~\ref{fig:cornerplot_gmos_uves} shows the posterior probability corner plot for solution UG1, demonstrating the credible value range for each orbital parameter. Fig.~\ref{fig:rvfit_gmos_uves} shows the radial-velocity curve in time and phase space produced from the MCMC outputs for solution UG1 compared to the observed data. The posterior probability plot and radial-velocity curve for solution U1 can be found in the Appendix~\ref{appen:solU1} (Fig.~\ref{fig:cornerplot_uves} \& Fig.~\ref{fig:rvfit_uves}). 
A summary of the associated orbital parameters for these two solutions can be found in Table~\ref{tab:orbital_parameters} and reveals that for most parameters they are reasonably consistent. Fig.~\ref{fig:xray_peri} shows the X-ray light curve from \citet{pol2018} using the solution UG1, confirming the X-Ray maximum is associated with the periastron of the orbit.

\begin{table*}
	\caption{Orbital parameters derived from radial-velocity fits using the combined VLT/UVES and Gemini/GMOS data (solution UG1) compared with solely the VLT/UVES data (solution U1) and the automated template fitting method (solution U2) discussed in Sect.~\ref{sec:auto_temp}. Quoted solutions represent the median of the posterior distributions from MCMC fitting. The error bars represent the 16th and 84th percentiles}
	\label{tab:orbital_parameters}
	\begin{tabular}{c@{\hspace{5mm}}c@{\hspace{5mm}}c@{\hspace{5mm}}c@{\hspace{5mm}}c@{\hspace{5mm}}c@{\hspace{5mm}}}
		\hline\hline
		          Parameter          &       VLT/UVES \&        &         VLT/UVES         &           \multicolumn{3}{c}{VLT/UVES}           \\
		                             &       Gemini/GMOS        &                          &         \multicolumn{3}{c}{Automated Fitting}\\
		                             &   \textit{Solution UG1}   &   \textit{Solution U1}   & \textit{N\,{\sc iv} 4058} & \textit{N\,{\sc v} 4945} & \textit{Solution U2} \\
        \hline
             $K_{\text{A}} $ [\kms]  & 130 $\pm$ 7              & 145 $\pm$ 10             & 132 $\pm$ 2               & 141 $\pm$ 6              & 137$\pm$ 3\\
	         $K_{\text{B}} $ [\kms]  & 141 $\pm$ 6              & 128 $\pm$ 9              & 126 $\pm$ 2               & 126 $\pm$ 4              & 127 $\pm$ 2\\
$\gamma$ velocity [\kms]             & 287 $\pm$ 5              & 277 $\pm$ 6              & 260 $\pm$ 1               & 283 $\pm$ 1              & 271 $\pm$ 1\\
		              e              & 0.68 $\pm$ 0.02          & 0.66 $\pm$ 0.03          & 0.764 $\pm$ 0.006         & 0.753 $\pm$ 0.011        & 0.758 $\pm$ 0.006\\
		$ \omega $ [$\degree$]       & 20.9 $\pm$ 3.8           & 28.8 $\pm$ 6.6           & 36.6 $\pm$ 0.9            & 40.1 $\pm$ 1.4           & 38.4 $\pm$ 0.8\\
        $P_{\text{orb}} $ [days]     & 154.55 $\pm$ 0.05        & 154.5 $\pm$ 0.7          & 155.03 $\pm$ 0.07         & 155.19 $\pm$ 0.12        & 155.11 $\pm$ 0.07 \\
	        $T_{0} $ [MJD]           & 57671.2 $\pm$ 0.9        & 57672.4 $\pm$ 2.0        & 57670.6$\pm$ 0.2          & 57670.8 $\pm$ 0.4        &  57670.7 $\pm$ 0.2\\ 
                      q              & 0.92 $\pm$ 0.07          & 1.13 $\pm$ 0.10          & 1.05 $\pm$ 0.02           & 1.11 $\pm$ 0.04          & 1.08 $\pm$ 0.02 \\		
   M$_{\text{A}}\sin^{3}(i)$ [\msol] & 65 $\pm$ 7               & 64 $\pm$ 13              &           -               &              -           & 39 $\pm$ 2\\
   M$_{\text{B}}\sin^{3}(i)$ [\msol] & 60 $\pm$ 7               & 73 $\pm$ 15              &           -               &              -           & 42 $\pm$ 3\\
   $a$$\,\sin(i)$ [AU]               & 2.82 $\pm$ 0.09          & 2.91 $\pm$ 0.18          &           -               &              -           & 2.45 $\pm$ 0.04\\
		  \hline\hline
	\end{tabular}
\end{table*}

Using the orbital parameter outputs from the MCMC fitting we can infer the minimum mass estimates for each component based on our solutions (Table~\ref{tab:orbital_parameters}) with Eqn.~\ref{eqn:mass}. 

\begin{equation}
	\text{M}_{\text{A,B}}\sin^{3}(i) = \frac{\text{P}}{2 \pi G}\ (1-e^{2})^{1.5}\ (\text{K}_{\text{A}} + \text{K}_{\text{B}})^{2}\ \text{K}_{\text{B,A}}
	\label{eqn:mass}
\end{equation}

Solution UG1 reveals M$_{\text{A}}\sin^{3}(i) = 65 \pm 7$~\msol\ and M$_{\text{B}}\sin^{3}(i) = 60 \pm 7$~\msol, whereas solution U1 reveals slightly higher determinations of M$_{\text{A}}\sin^{3}(i) = 64 \pm 13$~\msol\ and M$_{\text{B}}\sin^{3}(i) = 73 \pm 15$~\msol. 
We also note the reversal of the mass ratio between solution UG1 and U1 which is caused by the 10~\kms shift in the systemic velocity between the two solutions. Minimum semi major axis values are $a$$\,\sin^{3}(i) = 2.82 \pm 0.09$~AU and $a$$\,\sin^{3}(i) = 2.91 \pm 0.18$~AU for solutions UG1,U1 respectively. 

Earlier photometry of the 30 Doradus region by \citet{mas2002b}, obtained over a $\sim$3 week period using HST/STIS, revealed evidence for photometric variations in Mk34. The study notes a repeating periodic dip of $\sim$0.1~mag in Mk34, following a 20 day period, which is unrelated to the orbital phase.
Searches for an optical eclipse using the UVES acquisition camera images and performing relative photometry, were unsuccessful. The variable seeing and crowded region meant that relative magnitude variations were within the photometric accuracy of $\sim$0.2~mag achievable with this data.


\subsection{Automated Template Fitting}
\label{sec:auto_temp}

An alternative method for deriving the orbital parameters using an automated template fitting MCMC model was also applied to the UVES data-set. Similar to the previous approach, the model parameters consisted of the Keplerian elements ($K_{\text{A}}$, $K_{\text{B}}$, $\gamma$, e, $\omega$, P and T$_{0}$), a continuum flux ratio parameter of the two stars and a line dilution factor for star B. Uninformative priors were used on all parameters, except the period, which was a Gaussian prior of 154.5 $\pm$ 0.5 days, based on solution UG1 and the results from the phase dispersed minimisation. Using these parameters, and the same WN5h template spectrum of VFTS 682 from \citet{bes2011, bes2014}, we produced two scaled and shifted versions of the template to represent star A and star B. We used these templates to fit the N\,{\sc iv} 4058 and N\,{\sc v} 4945 emission lines separately, however the fit was applied to all 26 epochs of observations simultaneously. We also repeat the fitting from a range of starting points to ensure the results converged.

\begin{figure*}
	\includegraphics[width=\textwidth]{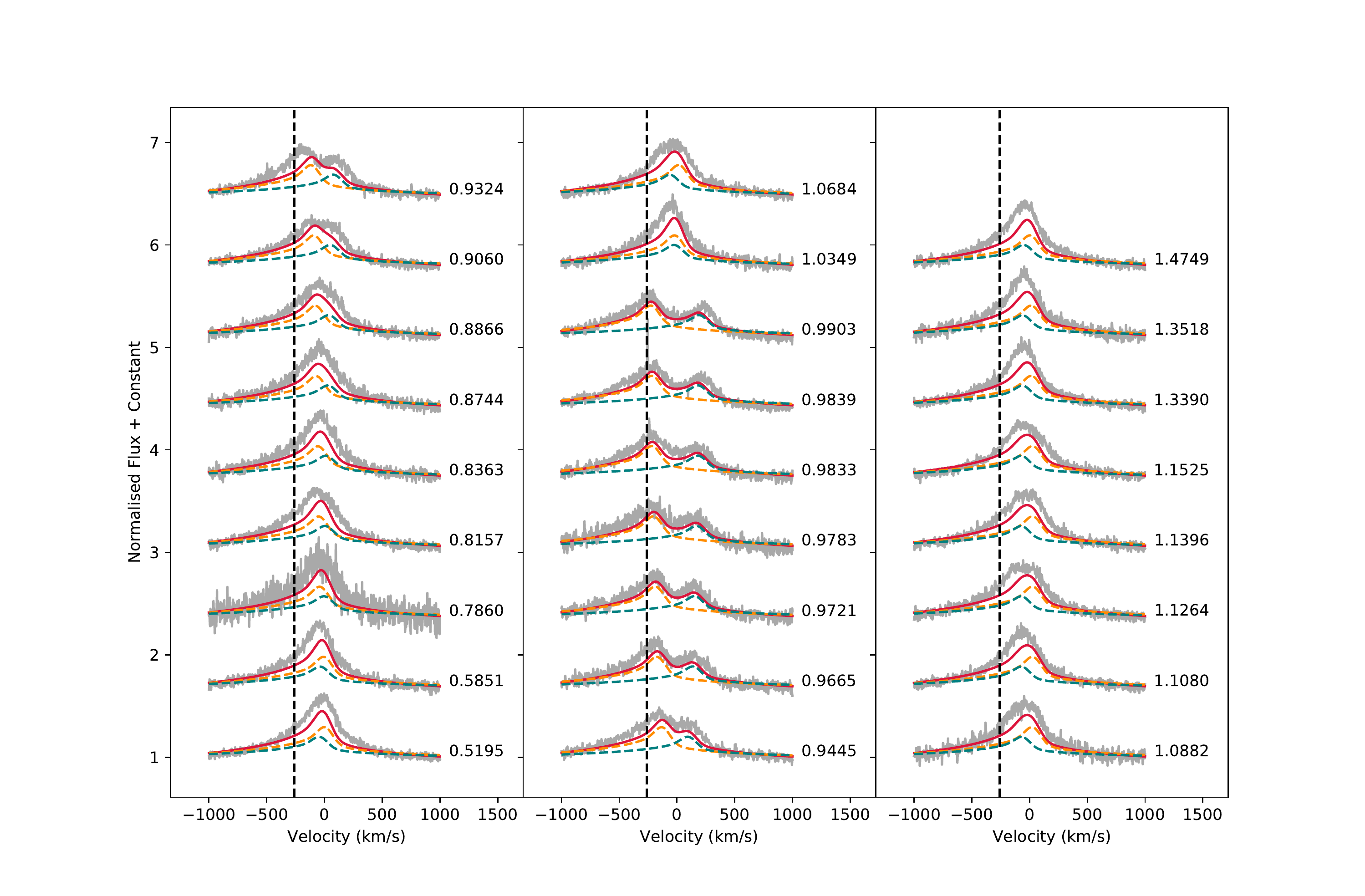}
	\caption{Automated template fits of the N\,{\sc iv} 4058 emission line for all epochs. The model for star A is shown in orange, star B in green and the combined fit is shown in red. The observed UVES data is shown in grey. The radial-velocity scale is centred such that the systemic velocity is zero and the dashed line corresponds to the rest wavelength of N\,{\sc iv} 4058.}
	\label{fig:auto_temp_niv}
\end{figure*}

\begin{figure*}
	\includegraphics[width=\textwidth]{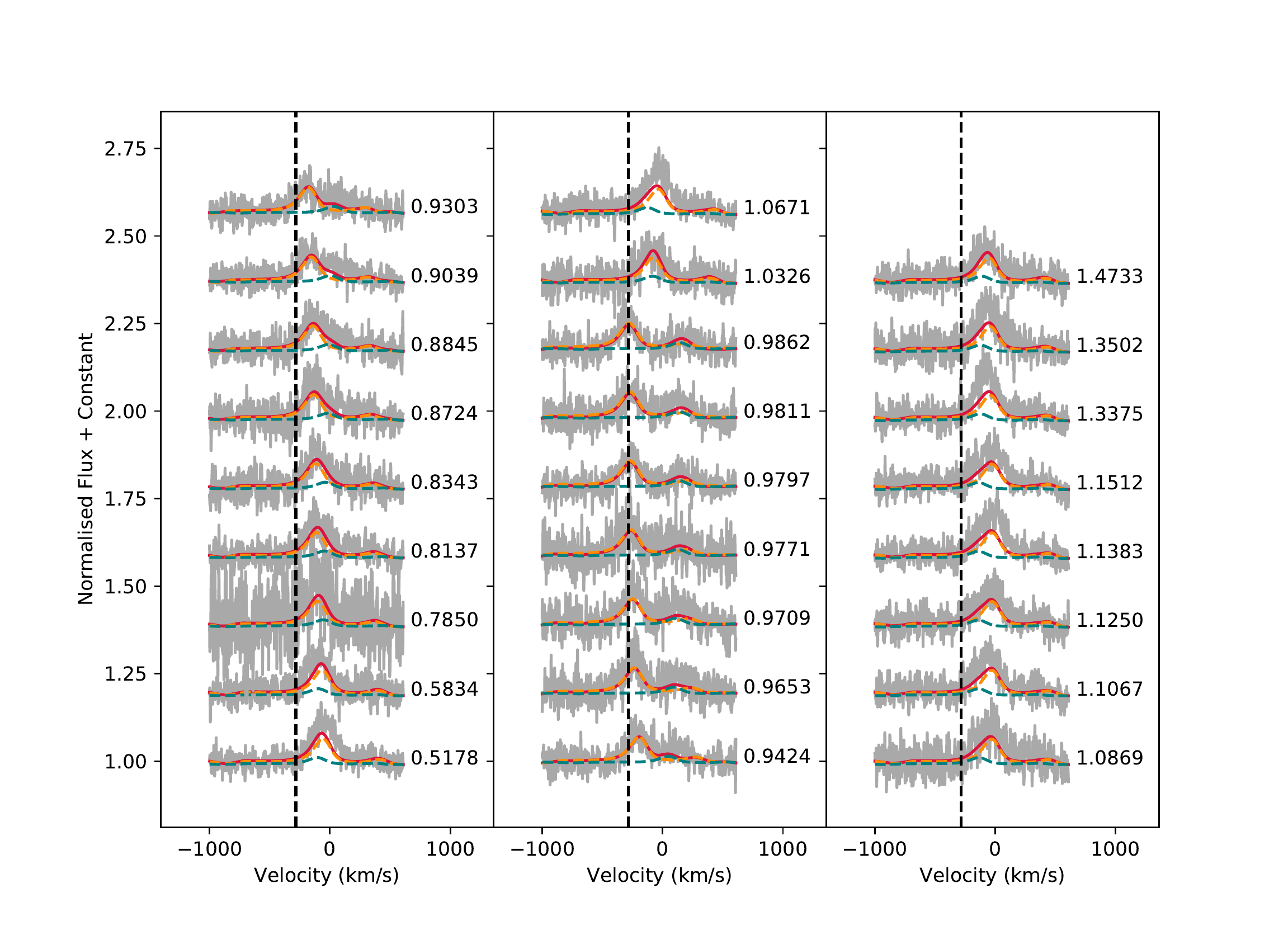}
	\caption{Automated template fits of the N\,{\sc v} 4945 emission line for all epochs. The model for star A is shown in orange, star B in green and the combined fit is shown in red. The observed UVES data is shown in grey. The radial-velocity scale is centred such that the systemic velocity is zero and the dashed line corresponds to the rest wavelength of N\,{\sc v} 4945.}
	\label{fig:auto_temp_nv}
\end{figure*}

Fig.~\ref{fig:auto_temp_niv} and Fig.~\ref{fig:auto_temp_nv} show the best fit models to the N\,{\sc iv} 4058 and N\,{\sc v} 4945 emission lines respectively. The probability distributions of the orbital parameters resulting from these fits are provided in Fig.~\ref{fig:cornerplot_auto} in Appendix~\ref{appen:solU2}. A summary of the best fit orbital solutions for each emission line is included in Table~\ref{tab:orbital_parameters} together with an average of both emission lines combined (solution U2).

\section{Physical and Wind Properties}
\label{sec:phys_and_wind}

\subsection{Spectral Analysis}
\label{sec:spec}

\begin{figure*}
	\includegraphics[width=0.85\textwidth]{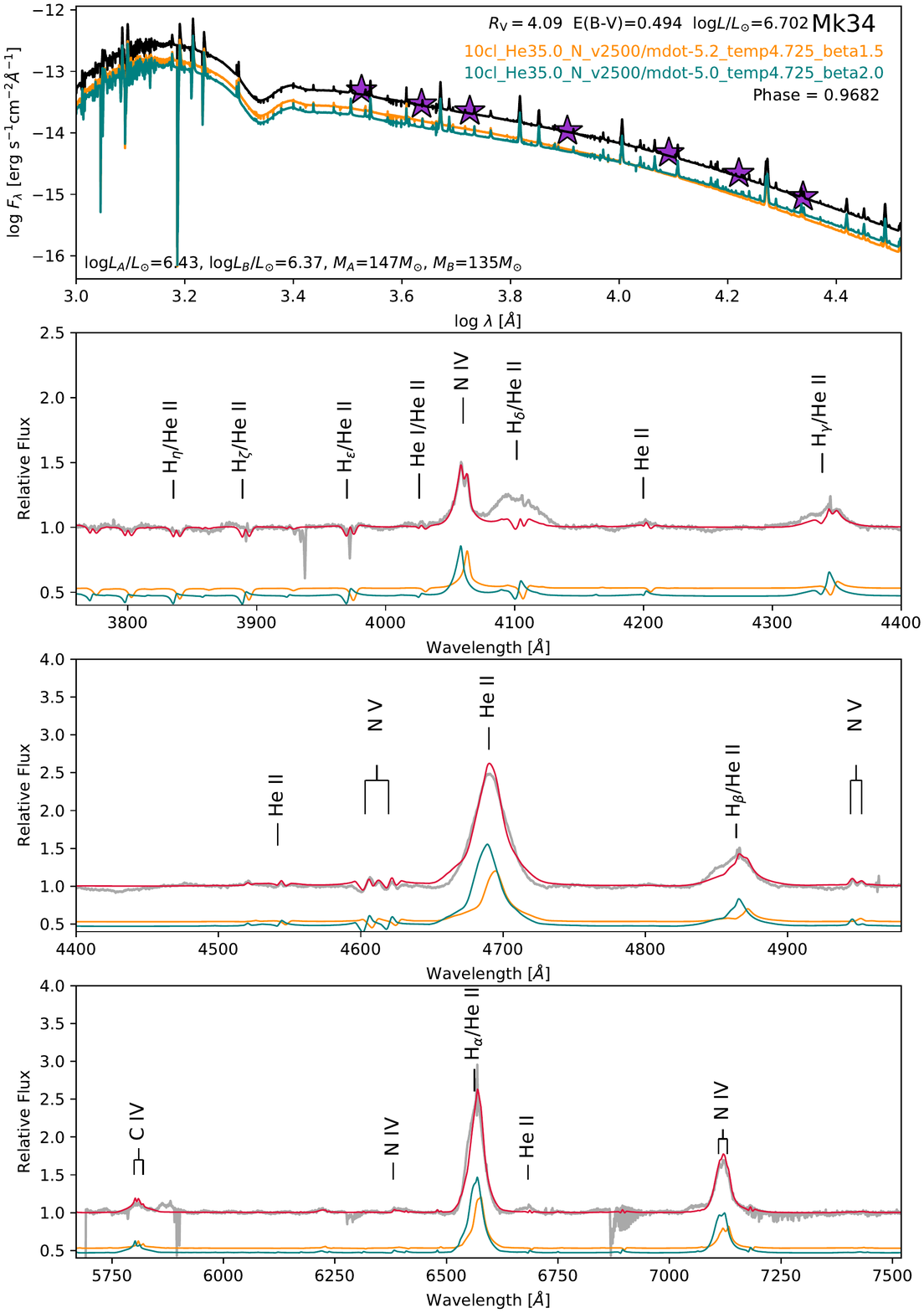}
	\caption{Top panel shows the individual reddening corrected model SED for star A (orange) and star B (green) produced by the radiative transfer code {\sc cmfgen}, and the combined SED for the system (black). The purple stars correspond to the observed magnitudes in various bands (listed in Table~\ref{tab:obs_mag}) used to derive the extinction towards the system. Lower panels show the observed VLT/UVES spectrum of epoch 2 (grey) with the chosen model for star A (orange), star B (green) and the combined model (red) over-plotted. Note the poor fit to the \hdelta and \hgamma emission is due to the observed excess in these lines close to periastron, as described in Sect.~\ref{sec:classification}}
	\label{fig:sed_model}
\end{figure*}

Both stellar components were identified as WN5h stars using the \citet{bes2011, bes2014} theoretical template spectrum for VFTS 682. Using the spectral properties of this star as a starting point we proceeded to compute a grid of models using the non-LTE stellar atmosphere and radiative transfer code {\sc cmfgen} \citep{hil1998}, varying effective temperature (at optical depth $\tau$=2/3), mass loss rates, helium mass fractions, terminal velocities and the $\beta$-type velocity law. For this template, we updated the N\,{\sc v} model atom better to reflect higher quantum number transitions\footnote{https://www.nist.gov/pml/atomic-spectra-database}. In particular, this caused a shift in the N\,{\sc v} 4945 emission line centroid of $\sim$26~\kms which would reduce the systematic offset between the measured radial velocities of the N\,{\sc iv} 4058 and N\,{\sc v} 4945 emission lines found in Sect.~\ref{sec:observations} such that the average differences would fall to 16~\kms and 7~\kms for UVES and GMOS respectively.
Using the mass ratio deduced from Solution UG1 (M$_{\text{B}}$/M$_{\text{A}}$ = q = 0.92 $\pm$ 0.07) and a specified He abundance, we estimate a luminosity ratio using a theoretical M/L relation (Eqn. 9 of \citealt{gra2011}) and based on this weighting we combine the two spectra and spectral energy distributions (SED) from each star. Each spectrum is shifted in accordance with the radial-velocity measurements recorded from the VLT/UVES data in Table~\ref{tab:OBs_uves_gmos} to reproduce the observed spectrum. An example of the combined spectrum fitted to the observed data, along with individual and combined SEDs, is shown in Fig.~\ref{fig:sed_model}.

Fits to the N\,{\sc iv} 4058, N\,{\sc v} 4603-20 P Cygni profile and N\,{\sc v} 4945 provided temperature estimates of 53,000 $\pm$ 1,200~K for both stars. The He\,{\sc ii} 4686 and \halpha\ emission line fits were used to deduce mass loss rates and He abundance mass fractions. For star A we find log ($\dot{\text{M}}_{\text{A}}$/[M$_{\odot}\text{yr}^{-1}]$) = $-$4.88 $\pm$ 0.13 and X$_{\text{He}}$ = 35 $\pm$ 5~\%. Similarly for star B we find log ($\dot{\text{M}}_{\text{B}}$/[M$_{\odot}\text{yr}^{-1}]$) = $-$4.72 $\pm$ 0.13 and X$_{\text{He}}$ = 35 $\pm$ 5~\%. We assume a nitrogen rich, CNO processed atmosphere at 0.5~\zsol\ metallicity, with a fixed nitrogen mass fraction of X$_{\text{N}}$ = 0.35~\%. This matched the observed nitrogen intensity and is equivalent to a $\sim$25x enhanced cosmic LMC nitrogen abundance. The volume filling factor was fixed at f$_{v}$ = 0.1 which is consistent with the electron scattering wings in the \halpha\ emission. The $\beta$-type velocity law was investigated by fitting the line profiles of the \hbeta\ and \hgamma\ lines and we find $\beta$ $\sim$1.5 and $\beta$ $\sim$2.0 for star A and star B respectively. The terminal velocity was estimated to be 2500 \kms for both components from fitting the line broadening of \halpha\ and He\,{\sc ii} 4686.
 
\begin{table}
	\caption{Mk34 photometry from the literature and the subsequently derived reddening parameters from this work.}
	\label{tab:obs_mag}
	\begin{tabular}{c@{\hspace{5mm}}c@{\hspace{5mm}}@{\hspace{5mm}}c}
		\hline\hline
		Parameter            & Value  & Reference \\ 
		\hline
		U [mag]              & 12.083            & \citet{dem2011} \\
		V [mag]              & 13.088            & \citet{dem2011} \\
		B [mag]              & 13.388            & \citet{dem2011} \\
		I [mag]              & 12.549            & \citet{dem2011} \\
		J [mag]              & 12.056            & \citet{skr2006} \\
		H [mag]              & 11.789            & \citet{skr2006} \\
		K$_{s}$ [mag]        & 11.68             & \citet{cro2010} \\
		R$_{\rm{V}}$ [mag]   & 4.09              & This work \\
		E(B-V) [mag]         & 0.49              & This work \\
		E(V-K$_{s}$) [mag]   & 1.76              & This work \\
		A$_{\rm{V}}$ [mag]   & 2.02              & This work \\
		A$_{\rm{K_s}}$ [mag] & 0.26              & This work \\
		DM [mag]             & 18.49             & \citet{pie2013} \\
		M$_{\rm{V}}$ [mag]   & $-$7.42           & This work \\
		M$_{\rm{K_s}}$ [mag] & $-$7.07           & This work \\
		\hline\hline
	\end{tabular}
\end{table}

The combined SED was used to examine the reddening and luminosity of the system. Intrinsic B, V and K-band colours were extracted by convolving the spectra with the corresponding filter function for each respective filter. Literature observed magnitudes outlined in Table~\ref{tab:obs_mag} were used in combination with these intrinsic magnitudes to derive E(B-V) and E(V-K$_{s}$) extinctions, which in turn were used with Eqn.~\ref{eqn:rv}, inferred from \citet{mai2014}, to derive a reddening parameter:

\begin{equation}
	R_{\rm{V}} = 1.2 \times \frac{\text{E(V-K}_{s}\text{)}}{\text{E(B-V)}} - 0.18
	\label{eqn:rv}
\end{equation}

We find R$_{V}$ = 4.09, which is typical of other stars in close proximity to R136 \citep{dor2013, gra2011}. The model spectrum was then reddened using the extinction law described in \citet{mai2014} along with our derived extinction and reddening parameters. Following this, we matched the observed K$_{s}$-band flux with the reddened model SED, whilst also accounting for the 50~kpc distance to the LMC \citep{pie2013}, to derive the absolute luminosity of each component in the system. 

Chemically-homogeneous masses\footnote{Conventionally, spectroscopic masses refer to masses derived from a surface gravity estimate. However, in the absence of a clean gravity diagnostic line, throughout this paper we refer to the chemically-homogeneous masses derived from the theoretical mass-luminosity relationship in this section as spectroscopic masses} were derived from the theoretical mass-luminosity relation from \citet{gra2011}, providing an estimate of the mass of each component. We find M$_{\text{A}}$ = 147 $\pm$ 22~\msol\ and M$_{\text{B}}$ = 136 $\pm$ 20~\msol\ for stars A and B, respectively. We consider the use of a mass-luminosity relationship appropriate since the low helium abundance of 35~\% is evidence that these stars are still on the main sequence. A summary of the spectral analysis results can be found in Table~\ref{tab:spec_evo_properties}.
 
\begin{table}
	\caption{Upper panel shows stellar properties derived from spectral modelling of the VLT/UVES spectra using the non-LTE stellar atmosphere code {\sc cmfgen}. Spectral masses were derived from the theoretical M/L relation from \citet{gra2011}. Lower panel shows stellar properties derived from evolutionary modelling of each component using the {\sc bonnsai} statistical analysis code. Luminosity, effective temperature and helium abundance inputs derived from spectral modelling were provided. All evolutionary parameters quote the mode result with the exception of He abundance which refers to the median.}
	\label{tab:spec_evo_properties}
	\begin{tabular}{c@{\hspace{5mm}}c@{\hspace{5mm}}c@{\hspace{5mm}}}
		\hline\hline

        \multicolumn{3}{c}{\textit{Spectroscopic modelling: {\sc cmfgen}}} \\
		Parameter & Star A & Star B \\\hline
		T [K]                                                         & 53000 $\pm$ 1200   & 53000 $\pm$ 1200 \\
		log (L$_{\text{*}}$/[L$_{\odot}]$)                            & 6.43 $\pm$ 0.08    & 6.37 $\pm$ 0.08 \\
		R [\rsol]                                                     & 19.3 $\pm$ 2.8     & 18.2 $\pm$ 2.7 \\
		log ($\dot{\text{M}}_{\text{*}}$/[M$_{\odot}\text{yr}^{-1}]$) & $-$4.88 $\pm$ 0.13 & $-$4.72 $\pm$ 0.13 \\
		v$_{\infty}$ [\kms]                                           & 2500 $\pm$ 300     & 2500 $\pm$ 300 \\
        $\beta$ (Velocity Law)                                        & 1.5                & 2.0 \\
		X$_{\text{He}}$ [\%]                                          & 35 $\pm$ 5         & 35 $\pm$ 5 \\
 		M$_{\text{cur}}$ [\msol]                                      & 147 $\pm$ 22       & 136 $\pm$ 20 \\
		\hline\hline	

        \multicolumn{3}{c}{\textit{Evolutionary modelling: {\sc bonnsai}}} \\
		Parameter & Star A & Star B \\\hline
		log (L$_{\text{*}}$/[L$_{\odot}]$)                                & 6.41$ ^{+ 0.09} _{-0.08}$    &  6.35$ ^{+ 0.08} _{-0.09}$   \\
		X$_{\text{He}}$ [\%]                                              & 33$ ^{+ 3} _{-8}$            &   33$ ^{+ 3} _{-8}$ \\
		v$_{\text{rot}}$ [\kms]                                           & 240$ ^{+ 171} _{-20}$        &  250$ ^{+ 170} _{-29}$ \\
		T [K]                                                             &  54388$ ^{+ 327} _{-822}$    & 54355$ ^{+ 339} _{-855}$  \\
		log ($\dot{\text{M}}_{\text{*}}$/[M$_{\odot}\text{yr}^{-1}]$)     & $-$5.00$ ^{+ 0.13} _{-0.11}$ & $-$5.06$ ^{+ 0.11} _{-0.12}$ \\
		Age [Myrs]                                                        &  0.5 $\pm$ 0.3               &  0.6 $\pm$ 0.3 \\
		M$_{\text{cur}}$ [\msol]                                          & 139$ ^{+ 21} _{-18}$         & 127$ ^{+ 17} _{-17}$   \\
		M$_{\text{init}}$ [\msol]                                         & 144$ ^{+ 22} _{-17}$         & 131$ ^{+ 18} _{-16}$  \\
		\hline\hline

	\end{tabular}
\end{table}


\subsection{Evolutionary Modelling}
\label{sec:evo}

Using the {\sc bonnsai}\footnote{The {\sc bonnsai} web-service is available at www.astro.uni-bonn.de/stars/bonnsai} evolutionary modelling code \citep{sch2014} we also derive evolutionary masses for Mk34 based on LMC stellar models \citep{bro2011, koh2015}. Providing luminosity, effective temperature and helium abundances derived from the spectral analysis the evolutionary modelling returned the parameters summarized in Table~\ref{tab:spec_evo_properties}. 

We find some differences between spectral and evolutionary modelling but attribute these to the mass-loss prescription adopted in the {\sc bonnsai} analysis, which under-predicts our observed mass-loss rates. To attain the high He abundance selected as a prior, the model requires high rotational velocity rates to enhance the He abundance at the surface of the star. Consequently, the high rotational velocity gives the star a higher effective temperature than expected. 
In general we cannot confirm the high rotational velocities predicted by the model. We can use the N\,{\sc iv} 4058, N\,{\sc v} 4603--20 and N\,{\sc v} 4945 emission to set good constraints on this parameter, and we find rotational broadening of $\sim$125~\kms, we cannot discern if this broadening is purely rotational or contaminated by micro and macro turbulence.

The {\sc bonnsai} code also provides estimates for the ages and initial masses of the stars. We find consistent stellar ages of $\sim$0.6 $\pm$ 0.3~Myrs for both stars, and current masses of M$_{\text{A(cur)}}$ = 139$ ^{+ 21} _{-18}$~\msol, and M$_{\text{B(cur)}}$ = 127$ ^{+ 17} _{-17}$~\msol\ for star A and star B respectively. These are somewhat lower than the spectral analysis results, which were based on the \citet{gra2011} mass-luminosity relation for chemical homogeneous stars. The initial masses corresponding to the BONNSAI solution are M$_{\text{A(init)}}$ = 144$ ^{+ 22} _{-17}$~\msol, and M$_{\text{B(init)}}$ = 131$ ^{+ 18} _{-16}$~\msol\ for star A and star B respectively.


\section{Discussion}
\label{sec:discussion}

\subsection{Melnick 34}
\label{sec:mk34}

In Sect.~\ref{sec:orb_sol} we have presented three different orbital solutions for Melnick 34 using two alternative methods from which we can confidently conclude that this is an eccentric, massive system with a mass ratio close to unity.

Solutions UG1 and U1 were derived using the same method and the results show good agreement indicating the method is robust. Similarly, solutions U1 and U2 were derived using the same data, though different methods, and again the results are comparable with the exception of eccentricity. Solution U2 obtains e = 0.758 $\pm$ 0.006, somewhat higher than the e = 0.66 $ \pm $ 0.03 found for solution U1. Whilst there are advantages to the automated model fitting, such as the unbiased and systematic nature of the method, we were unable to apply this technique to the GMOS data due to its low S/N ratio. We therefore favour solution UG1 which encompasses all available data, appropriately weighted.

Using these results we can comment on the inclination of the system, by comparing the predicted dynamical masses with the spectroscopic masses derived in Sect.~\ref{sec:spec}. Table~\ref{tab:inc} shows how the kinematic masses and periastron distances depend on inclination for solutions UG1 and U2. Recalling the spectroscopic masses of M$_{\text{A}}$ = 147 $\pm$ 22~\msol\ and M$_{\text{B}}$ = 136 $\pm$ 20~\msol, it is apparent from Table~\ref{tab:inc} that solution UG1, with an eccentricity of 0.68, would reproduce the dynamical masses if i $\sim$50$\degree$. In contrast, solution U2, with a higher eccentricity of 0.758, suggests i $\sim$41$\degree$ to reconcile the dynamical and spectroscopic masses.

\begin{table}
	\caption{A comparison between the stellar masses and periastron distances for Mk34 for various assumed inclination values for both Solution UG1 and Solution U2 orbital properties.}
	\label{tab:inc}
	
	\begin{tabular}{c@{\hspace{5mm}}c@{\hspace{5mm}}c@{\hspace{5mm}}c@{\hspace{5mm}}c@{\hspace{5mm}}c@{\hspace{5mm}}c@{\hspace{5mm}}}
		\hline \hline
		 & \multicolumn{3}{c}{\textit{Solution UG1}}                                    & \multicolumn{3}{c}{\textit{Solution U2}} \\
		 Incl                  & M$_{\text{A}}$ & M$_{\text{B}}$    & a$_{\text{peri}}$ & M$_{\text{A}}$ & M$_{\text{B}}$ & a$_{\text{peri}}$ \\
		                       & [\msol]        & [\msol]           & [AU]              & [\msol]        & [\msol]        & [AU] \\
		 \hline
		 80$\degree$           & 68             & 63                & 0.91              & 41             & 44             & 0.60 \\
		 70$\degree$           & 78             & 72                & 0.96              & 47             & 51             & 0.63 \\
		 60$\degree$           & 100            & 92                & 1.04              & 60             & 65             & 0.68 \\
		 50$\degree$           & 144            & 133               & 1.18              & 87             & 94             & 0.77 \\
		 40$\degree$           & 244            & 225               & 1.40              & 147            & 159            & 0.92 \\
		 \hline \hline
	\end{tabular}	
\end{table}

We are in the process of investigating the inclination of the system using X-ray light curve modelling. Another approach would be to search for a visual eclipse, however this would involve obtaining further, very high cadence, optical observations precisely focused on conjunction. Based on solution UG1, the lowest inclination that would produce an eclipse is $\sim$79$\degree$, from which we anticipate the duration of this eclipse to be approximately 16 hours. An inclination of $\sim$79$\degree$ would reveal dynamical masses of M$_{\text{A}}$ = 63~\msol\ and M$_{\text{B}}$ = 69~\msol, much lower than our spectral modelling estimates, therefore we consider an eclipse to be highly unlikely. Other observations necessary to help constrain the inclination include intensive X-ray monitoring near periastron, which we could combine with Fig.~\ref{fig:xray_peri} and allow us to accurately map, and subsequently model, the X-ray variability. Finally, polarimetry observations have proven useful to help measure the inclination of massive binaries, as done by \citet{stl1993} for WR139.


\subsection{Colliding Wind Binaries}
\label{sec:cwb}

Table~\ref{tab:comp} provides a comparison between the stellar and orbital properties derived here for Mk34 and other Wolf-Rayet binary systems with a >75~\msol\ primary mass in the Milky Way and LMC. Spectroscopic masses, evolutionary masses and exceptional L$_{\text{x}}$/L$_{\text{bol}}$ confirms that Mk34 is a high mass colliding wind binary and most likely the most massive binary known to date. Mk34 is also an excellent candidate for studying the properties of colliding wind binaries (CWB). Of the systems listed in Table~\ref{tab:comp} only WR21a and WR25 are CWB, making Mk34 a welcome addition at low metallicity, and WR25 is a SB1 system making mass derivations of the secondary component very uncertain.

Other CWBs outside this group include $\eta$ Carinae. Again the suitability of this system as the focus of detailed CWB studies is not ideal since the nature of the secondary is uncertain. The SB2 nature of Mk34 has allowed us to thoroughly investigate the orbital properties of this system and produce robust results. The high eccentricity and period encourage obvious wind variations throughout the orbit, which is also useful for studying CWB properties.

\citet{pol2018} show a Chandra ACIS-I X-ray image of Mk34 compared with neighbouring stars R136a1--3 and R136c (their Fig. 1). It has been proposed that R136a1--3 host stellar masses of 265~\msol, 195~\msol, and 135~\msol\ respectively \citep{cro2010} which in turn has sparked discussion regarding the possible binarity of these systems, with a mass ratios close to unity. In the binary scenario, R136a1 would be analogous to Mk34 and therefore we would also expect to observe similar X-ray properties if its orbital period were on the order of $\sim$100 days. Instead we find Mk34 is considerably brighter than the entire R136a cluster at all times \citep{tow2006} due to the colliding wind nature of the system. The lower X-ray flux of R136a support the interpretation that these stars are single, possess high mass ratios, or large separations for mass ratios close to unity.

\begin{table*}
	\begin{center}
	\caption{Comparison of stellar and orbital properties of other known massive binary systems in the Large Magellanic Cloud (LMC) and Milky Way (MW).}
	\label{tab:comp}
	\begin{tabular}{c@{\hspace{0.5mm}}c@{\hspace{0.5mm}}c@{\hspace{0.5mm}}c@{\hspace{0.5mm}}|c@{\hspace{0.5mm}}c@{\hspace{0.5mm}}c@{\hspace{0.5mm}}c@{\hspace{0.5mm}}c@{\hspace{0.5mm}}}
	\hline\hline
                                           & \multicolumn{3}{c}{LMC} &  \multicolumn{5}{c}{MW} \\
        \cmidrule(lr){2-4}\cmidrule(lr){5-9}
	Property                           & Melnick 34               & R145                & R144            &   NGC3603-A1   &      F2          &       WR 20a           &        WR 21a       & WR 25 \\
	Ref                                &           [1]            &         [5]         &       [8]       &      [9]       &      [14]        &          [15]          &   [19]              & [21]  \\ 
	\hline
	Spectral Type A                    & WN5h                     & WN6h                & WN5--6h          & WN6ha          & WN8--9h           & WN6ha$^{[16]}$         & O3/WN5ha              & O2.5 If*/WN6$^{[22]}$ \\
	Spectral Type B                    & WN5h                     & O3.5 If*/WN7        & WN6--7h          & WN6ha:         & O5--6 Ia          & WN6ha$^{[16]}$         & O3Vz((f*))             & O2.5 If*/WN6$^{[22]}$ \\
	Period [days]                      & 154.55 $\pm$ 0.05 &       158.76        &      <370       &       3.77     &       10.5       &3.69 $ \pm $ 0.01$^{[17]}$& 31.67 $\pm$ 0.01  & 207.85 $\pm$ 0.02  \\   
	Eccentricity                       &    0.68 $ \pm $ 0.02     &  0.788 $\pm$ 0.007  &        -        &       0        & 0.05 $\pm$ 0.01  &        0               & 0.695 $\pm$ 0.005   &  0.50 $\pm$ 0.02    \\
	Mass Ratio                         &  0.92 $\pm$ 0.07         &  1.01 $\pm$ 0.07  & 1.17 $\pm$ 0.06 &     0.76       & 0.73 $\pm$ 0.07  &       0.99             & 0.563 $\pm$ 0.025   &  - \\
	M$_{\text{A}}$sin$^{3}$(i) [\msol] & 65 $\pm$ 7               &   13.2 $\pm$ 1.9    &       -         &  99 $\pm$ 34   &   64 $\pm$ 9     &   74.0 $\pm$ 4.2       &  64.4 $\pm$ 4.8     &  - \\
	M$_{\text{B}}$sin$^{3}$(i) [\msol] & 60 $\pm$ 7               &   13.4 $\pm$ 1.9    &       -         &  75 $\pm$ 21   &   47 $\pm$ 6     &  73.3 $\pm$ 4.2        &  36.3 $\pm$ 1.7     &  - \\
	Inclination [$\degree$]            &            -             &     39 $\pm$ 6      &       -         &      71        &  67 $\pm$ 1      & 74.5 $\pm$ 1.0$^{[17]}$&        -            &  - \\
	M$_{\text{A(orb)}}$ [\msol]        &            -             &  53$^{+40} _{-20}$  &       -         &  116 $\pm$ 31  &  82 $\pm$ 12     &     82.7 $\pm$ 5.5     &        -            &  - \\
	M$_{\text{B(orb)}}$ [\msol]        &            -             &  54$^{+40} _{-20}$  &       -         & 89 $\pm$ 16    &  60 $\pm$ 8      &    81.9 $\pm$ 5.5      &        -            &  - \\ 
	\hline
	Distance [kpc]                     &          50$^{[2]}$      &     50$^{[2]}$      &   50$^{[2]}$    & 7 $\pm$ 0.5$^{[10]}$ &    8.1        & 8.0 $\pm$ 1.4$^{[18]}$ &   4.4$^{[20]}$      &  2.3$^{[23]}$\\
	M$_{\text{A(sp)}}$ [\msol]         &       147 $\pm$ 22       & 101$^{+40} _{-30}$  &       -         & 120$ ^{+26} _{-17}$$^{[11]}$ &   -  &   -            &        -            &  - \\
	M$_{\text{B(sp)}}$ [\msol]         &       136 $\pm$ 20       & 109$^{+60} _{-40}$  &       -         & 92$ ^{+16} _{-15}$$^{[11]}$ &   -  & -             &        -            &  - \\
	L$_{\text{x}}$ [$\times$10$^{34}$ \ergs] &  1--32$^{[3]}$      &  0.6$^{[6]}$       &   1$^{[6]}$    &     1.7--4.8   $^{[6,12]}$      &    -   &  -            &  0.1--1.3$^{[3,20]}$    & 0.4--0.9$^{[3]}$ \\
	M$_{\text{v}}$ [mag]               &       $-$7.42$^{[1,4]}$       &    $-$7.2$^{[7]}$     &      $-$8.2       & $-$7.5$^{[13]}$  &      -           &$-$7.04 $\pm$ 0.25$^{[18]}$ &   $-$5.9$^{[19,20]}$            &  $-$7.3$^{[24]}$ \\ 
	\hline\hline
	\multicolumn{9}{l}{
		\begin{minipage}{\textwidth}~\\
		1. This work; 2. \citet{pie2013}; 3. \citet{pol2018}; 4. \citet{dem2011}; 5. \citet{she2017}; 6. Townsley (private communication); 7. \citet{sch2009}; 8. \citet{san2013b}; 9. \citet{sch2008}; 10. \citet{mof1983}; 11. \citet{cro2010} 12. \citet{tow2014}; 13. \citet{dri1995}; 14. \citet{loh2018}; 15. \citet{rau2005}; 16. \citet{rau2004}; 17. \citet{bon2004}; 18. \citet{rau2007}; 19. \citet{tra2016}; 20. Rate (private communication); 21. \citet{gam2006}; 22. \citet{cro2011}; 23. \citet{smi2006}; 24. \citet{ham2006} \\
The colon (:) indicates a highly uncertain measurement \\			
	\end{minipage}
	}\\ 
	\end{tabular}
	\end{center}
\end{table*}


\subsection{Formation}
\label{sec:formation}

Our spectroscopic analysis reveals both components of Mk34 are VMS, which prompts the question of how this system formed. Both components are classified as WN5h stars, which are very massive main sequence stars with intrinsically powerful stellar winds. Mk34 lies at a projected distance of $\sim$2~pc from R136, the dense core of 30 Doradus, whose age is 1.5 $\pm$ 0.5~Myr \citep{cro2016}. The derived age of Mk34 ($\sim$0.6~Myr) would lead us to favour a scenario in which Mk34 was formed in situ rather than being ejected from this dense core. Proper-motion estimates of this system using Gaia DR2 suggest a $\sim$30 $\pm$ 15~\kms motion relative to R136 (D. Lennon, private communication). Given the systemic radial velocity of Mk34 from solution UG1, 287 $\pm$ 5~\kms, and the cluster radial velocity of R136, 268 $\pm$ 6~\kms \citep{hen2012}, we find a relative radial velocity offset of 19 $\pm$ 8~\kms, and therefore a relative 3D velocity of 36 $\pm$ 26~\kms. Considering this relative 3D velocity and the young age of Mk34, the system has traversed 22 $\pm$ 19~pc since birth, which exceeds the 2~pc projected distance so is likely inconsistent with a birth environment of R136, prefering a birth in isolation. Here we discuss two potential scenarios from which these massive stars could form: 

\begin{enumerate}
\item Core fragmentation;
\item Merger of smaller stars.
\end{enumerate}

The core fragmentation mechanism predicts that an initial, bound, proto-stellar core will fragment and collapse to form multiple stars \citep{goo2007}. From this we would expect a small group of stars would form, with a range of stellar masses. Current observations are unable to test a hypothetical low mass population around Mk34. It is also possible, given the close proximity of Mk34 to R136, that the radiation field from stars in R136 heated the neighbouring star formation regions and suppressed the fragmentation of low mass counterparts, instead allowing only the most massive stars to form \citep{dal2012, dal2013}. 

The alternative scenario involving merging stars begins in a similar manner to the core fragmentation hypothesis, however rather than the direct formation to the two massive stars we observe today, the cloud instead collapses to form a system with four or more massive stars, each with $\sim$70~\msol, which in time undergo mergers to form the current system. Again, this scenario would require an additional low mass population to be present in order to induce the hardening of the quadruple, thereby triggering the mergers. Alternatively, this hardening could be achieved through dynamical friction, where the surrounding ambient gas applies a braking force to the system and the loss of angular momentum causes the migration of the components inwards, resulting in a merger \citep{sta2010}.

To further investigate the likelihood of the merging scenario compared to core fragmentation limited by a background radiation field, a Monte Carlo simulation was used to explore the relative likelihood of forming exactly two $\sim$130 $\pm$ 30~\msol\ stars compared to four $\sim$70 $\pm$ 30~\msol\ stars, providing a rough insight into the more likely formation mechanism. Using a cluster mass function with a slope of $-$2, across an initial cluster mass range of 50--10,000~\msol, the simulation randomly sampled the cluster mass function and created a population of stars within each cluster by randomly sampling a \citet{mas2013b} IMF, with an initial mass range of 0.1--300~\msol. In total, 3.76 million clusters were simulated. For full details of this method see \citet{par2007}.

\begin{figure}
	\includegraphics[width=\columnwidth]{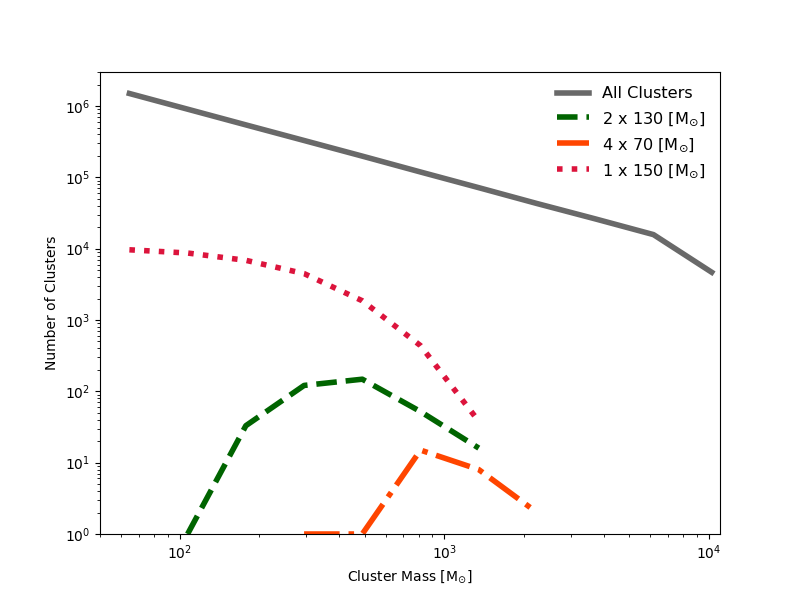}
	\caption{The results of the Monte Carlo simulation showing the number of clusters hosting either two $\sim$130~\msol\ stars (green dashed line), four $\sim$70~\msol\ stars (orange dot-dashed line), or 1 $\sim$150~\msol\ star (red dotted line). The likelihood of forming a cluster hosting 2 VMS is slightly higher than forming a cluster of 4 massive stars.}
	\label{fig:MC}
\end{figure}

By counting the number of clusters which successfully formed either exactly two $\sim$130 $\pm$ 30~\msol\ stars or four $\sim$70 $\pm$ 30~\msol\ stars, and no other O-stars ($\geq$17.5~\msol) we find that both scenarios are unlikely. The formation of two VMS, however, is more favourable than the formation of four massive stars, with probabilities of 0.01~\% and 0.0005~\% respectively. Fig.~\ref{fig:MC} shows a comparison of the results we obtained, showing more clusters forming two VMS rather than four massive stars. We also note that the host cluster masses from which the VMS form are between 420--1200~\msol. From this we are inclined to conclude that this system is most likely to have formed in-situ, through the direct formation of a low mass cluster including two VMS.

\subsection{Future Evolution}
\label{sec:future}

At present the components of Mk34 orbit in relatively close proximity, with a minimum periastron separation of $\sim$0.9~AU, however within the next 2-3~Myrs each component will evolve off the main sequence, changing the dynamics of the system. To predict these changes we compare the evolutionary model grids from the Bonn \citep{koh2015} and Geneva \citep{yus2013} codes, based on a star with an initial mass of 150~\msol\ at LMC metallicity (Z=0.006 and Z=0.0047 respectively) and a selected rotational-velocity. 

\begin{figure}
	\includegraphics[width=\columnwidth]{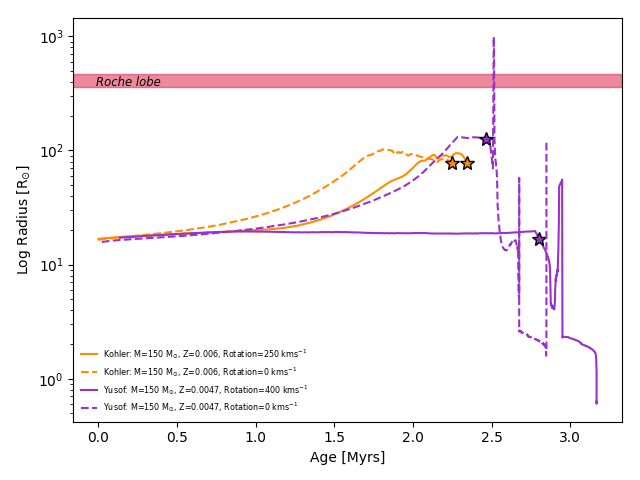}
	\caption{The evolution of the stellar radius with time for a 150~\msol star at LMC metallicity, derived from the \citet{koh2015} model grids (orange lines) for a star with 250~\kms rotational velocity (solid line) and a star with no rotation (dashed line). Also included are the \citet{yus2013} model grids (purple lines) for a star with a 400~\kms rotational velocity (solid line) and no rotation (dashed line). The red shaded region represents the Roche lobe radius range of the system between the zero age main sequence and terminal age main sequence, based on an inclination of 50\degree.}
	\label{fig:radius_time} 
\end{figure}

We begin with the model grids by \citet{koh2015} which describe the evolution of the star up until core hydrogen burning ceases, which is taken to be when the hydrogen core mass fraction drops below 3\%. These models were used in the {\sc bonnsai} evolutionary modelling in Sect.~\ref{sec:evo} for which we derived an initial rotational velocity of 250~\kms, therefore continuing with this particular model suggests the end of the hydrogen burning stage will occur after 2.3~Myrs. During this time the effective temperature of the star will fall to $\sim$29~kK, causing the radius to swell, reaching $\sim$78~\rsol. Fig.~\ref{fig:radius_time} shows how the stellar radius evolves through time and can be used to show if Roche lobe overflow and mass transfer in the system would be likely. 

For an inclination of 50\degree\ we find the current periastron separation to be 254~\rsol\ and using Eqn.~\ref{eqn:roche} \citep{egg2006} we find a Roche lobe limit of 386~\rsol. Considering that the periastron distance will change through time, and assuming a non-conservative, fast wind mass-loss mode as parametarised by \citet{pos2014}, we can derive a Roche lobe limit range based on the initial and final periastron separations at the zero age main-sequence and terminal age main-sequence times respectively. This range corresponds to the shaded region in Fig.~\ref{fig:radius_time}, and therefore we can see that based on the 250~\kms rotation \citet{koh2015} model grid the primary will not fill its Roche lobe at the end of core hydrogen burning.

\begin{equation}
	R_{\rm{L}} = a\ \frac{0.49\text{q}^{\nicefrac{2}{3}} + 0.27\text{q} - 0.12\text{q}^{\nicefrac{4}{3}}}{0.6\text{q}^{\nicefrac{2}{3}} + \ln(1 + \text{q}^{\nicefrac{1}{3}})}
	\label{eqn:roche}
\end{equation}

Assuming no initial rotation, the \citet{koh2015} models predict a very similar evolution, with a core-hydrogen burning lifetime of 2.2~Myrs, and an effective temperature of $\sim$28~kK at the terminal age main sequence, therefore deriving a radius of $\sim$78~\rsol. Again we can see in Fig.~\ref{fig:radius_time} that an interaction between the two stars is unlikely prior to their post main sequence evolution.

Since the \citet{koh2015} model grids terminate at the end of core hydrogen-burning, to understand how the star could progress post-hydrogen core burning we can use the model grids by \citet{yus2013} with two rotational-velocity options; no rotation and a rotational velocity of 404~\kms.

For single stars in isolation, \citet{yus2013} predict that at the end of the core hydrogen burning stage a 150~\msol\ initial mass star at LMC metallicity will have a mass of 66--72~\msol, depending on rotation. For the model grid with no rotation, hydrogen burning will cease after 2.5~Myrs, during which the effective temperature will have fallen to $\sim$22~kK, with a radius of $\sim$126~\rsol. Again, the evolution of the stellar radius through time is shown in Fig.~\ref{fig:radius_time} and it can been seen the expansion does not reach the Roche-lobe limit. Post core hydrogen-burning, there is a brief period of further expansion as can be seen in Fig.~\ref{fig:radius_time}, such that the star will fill its Roche-lobe, triggering Roche-lobe overflow and mass transfer onto the secondary. The alternative model grid introduces a rotational velocity of 404~\kms, and within this scenario the core hydrogen burning phase ends after 2.8~Myrs. Under these initial conditions the effective temperature of the star rises to $\sim$64~kK, resulting in a reduction of the stellar radius to $\sim$17~\rsol, suggesting there will be no interaction between the two stars.



Overall it is unlikely that Roche lobe overflow occurs, in which case evolution will proceeds independently. \citet{yus2013} predict that at the end of the helium burning stage a 150~\msol\ initial mass star will have a core mass of 43$\pm$5~\msol, and that these stars will end their lives with the direct collapse to a black hole or black hole formation following a weak core collapse supernova. Unfortunately similar models for binary systems are not yet available, though see \citet{kru2018} and \citet{bel2016} for potential massive binary star progenitors of LIGO gravitational wave sources. Under the assumption of single stellar models applying to Mk34, and it surviving two potential supernovae events, this system will progress to be a double stellar-mass black hole binary with a combined mass of $\sim$90~\msol. Considering the current gravitational wave catalogue by \citet{ligo2018}, which reveals roughly equal mass BH-BH binary system progenitors, we find that our predicted evolution for Mk34 produces a system with components comparable to that of the gravitational wave source GW170729. We therefore conclude that the ultimate fate of Mk34 could be a merger event, albeit in the far distant future due to the large separation between these stars.


\section{Conclusion}
\label{sec:conclusion}

Our main results can be summarized as follows:

\begin{enumerate}

\item Using a combination of VLT/UVES and Gemini/GMOS spectroscopic monitoring we confirm Mk34 is an SB2 binary system, with closer inspection revealing both stars are of WN5h spectral type.

\item Radial-velocity measurements of the N\,{\sc iv} 4058 and N\,{\sc v} 4945 emission lines reveal a 154.5 day period, eccentric system (e = 0.68 $\pm$ 0.02) with a mass ratio of q = 0.92$\pm$0.07 as shown in solution UG1. Minimum masses of M$_{\text{A}}$sin$^{3}$(i) = 65 $\pm$ 7~\msol\ and M$_{\text{B}}$sin$^{3}$(i) = 60 $\pm$ 7~\msol\ were derived for star A and star B respectively.

\item Solution U2 used automated template fitting to the N\,{\sc iv} 4058 and N\,{\sc v} 4945 emission lines, which produces similar results. Using an average from both emission lines we find a 155.11 day period, an eccentricity of 0.758 $\pm$ 0.006 and a mass ratio of 1.08 $\pm$ 0.02, which gives minimum masses of M$_{\text{A}}$sin$^{3}$(i) = 39 $\pm$ 2~\msol\ and M$_{\text{B}}$sin$^{3}$(i) = 42 $\pm$ 3~\msol.

\item Spectral modelling using the template spectrum of VFTS 682 unveiled spectroscopic masses of 147 $\pm$ 22~\msol\ and 136 $\pm$ 20~\msol\ for star A and B respectively. Evolutionary modelling predicted slightly lower mass estimates of 139$ ^{+ 21} _{-18}$~\msol\ and 127$ ^{+ 17} _{-17}$~\msol\ for star A and star B respectively. For consistent dynamical and spectroscopic masses, the inclination of the system would need to be $\sim$50$\degree$, which is currently being investigated through X-ray light curve modelling (Russell et al. in prep).

\item Mk34 is an excellent addition to the CWB catalogue, with a high eccentricity producing significant wind variations and reasonable period length allowing detailed monitoring across the full orbit. The predicted masses suggest Mk34 is the highest mass binary system known, and further investigation into the inclination is worthwhile to verify this. 

\item Based on an initial rotational velocity of 250~\kms it is unlikely that in the future these stars will fill their Roche lobes and undergo mass transfer. Both components are, however, likely to evolve through to stellar mass black holes and therefore, should the binary survive, Mk34 will become a double stellar-mass black hole binary and a potential LIGO BH-BH source.

\end{enumerate}


\section*{Acknowledgements}

We are very grateful for the support provided by John Pritchard from the User Support department at ESO, especially for his solid advice and continued patience whilst acquiring the UVES data. We are also very grateful to our referee whose comments and feedback were invaluable to the analysis and interpretation of our results. We would like to thank Steven Parsons for his guidance using the REFLEX reduction pipeline, and Heloise Stevance \& Martin Dyer for sharing their time and python coding skills. Finally, we thank Patrick Broos and Leisa Townsley for their X-ray expertise and Gemma Rate for sharing her Gaia DR2 knowledge. KAT and JMB would like to thank STFC for financial support through the STFC Doctoral Training Grant and STFC consolidated grant (reference ST/M001350/1) respectively.  RJP acknowledges support from the Royal Society in the form of a Dorothy Hodgkin Fellowship.

Based on observations obtained at the Gemini Observatory acquired through the Gemini Observatory Archive, which is operated by the Association of Universities for Research in Astronomy, Inc., under a cooperative agreement with the NSF on behalf of the Gemini partnership: the National Science Foundation (United States), the National Research Council (Canada), CONICYT (Chile), Ministerio de Ciencia, Tecnolog\'{i}a e Innovaci\'{o}n Productiva (Argentina), and Minist\'{e}rio da Ci\^{e}ncia, Tecnologia e Inova\c{c}\~{a}o (Brazil).
This publication makes use of data products from the Two Micron All Sky Survey, which is a joint project of the University of Massachusetts and the Infrared Processing and Analysis Center/California Institute of Technology, funded by the National Aeronautics and Space Administration and the National Science Foundation.


\bibliographystyle{mnras}


\appendix

\section{Orbital Properties -- Solution U1}
\label{appen:solU1}
\begin{figure*}
	\includegraphics[width=0.7\textwidth]{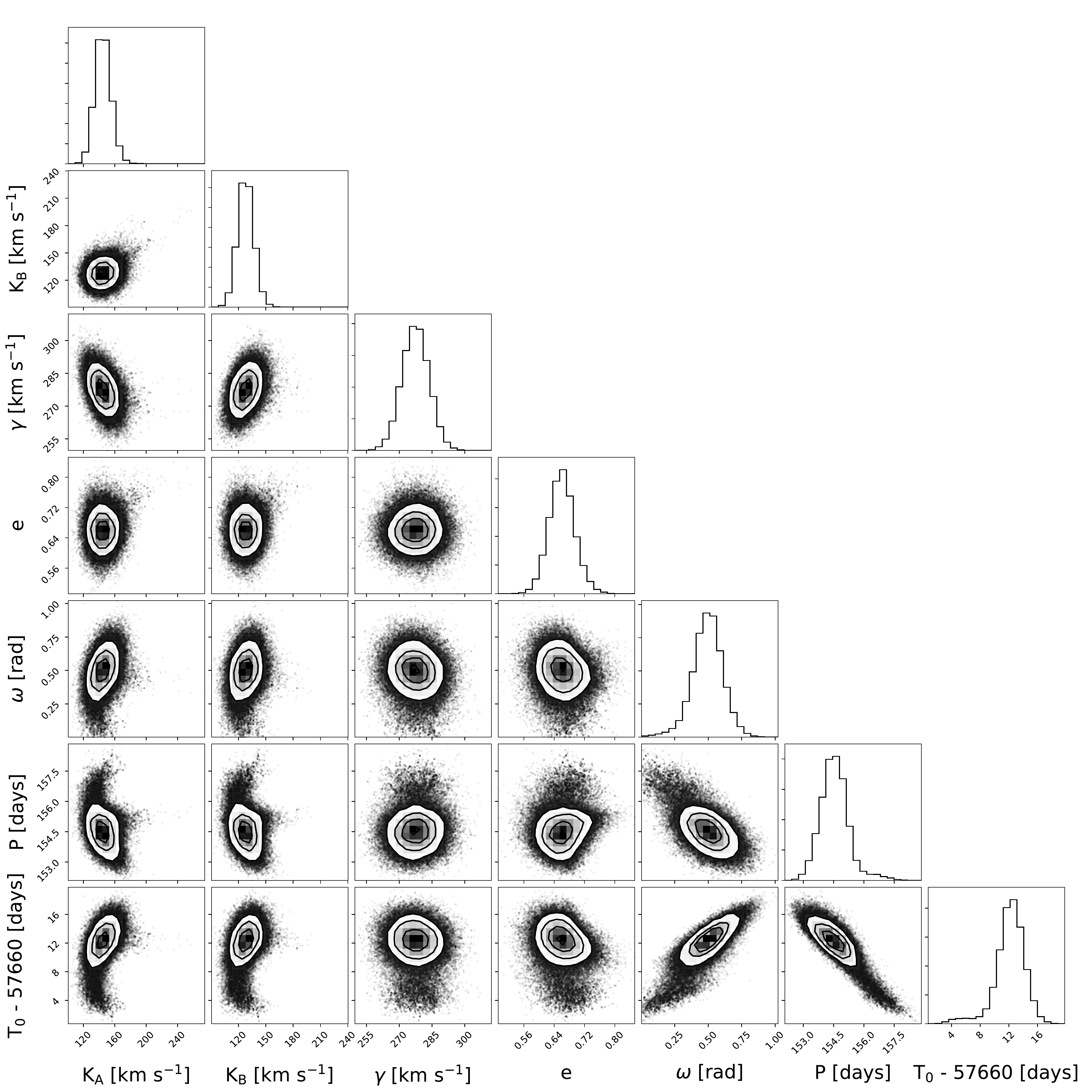}
	\caption{Corner plot showing posterior probabilities for solution U1, where $K_{\text{A}}$ and $K_{\text{B}}$ are the semi-amplitudes of the velocities for star A and star B respectively, $\gamma$ is the systemic velocity, e is the eccentricity, omega is the longitude of the periastron, P is the orbital period and T$_{0}$ is the time of periastron.}
	\label{fig:cornerplot_uves}
\end{figure*}


\begin{figure*}
	\includegraphics[width=0.7\textwidth]{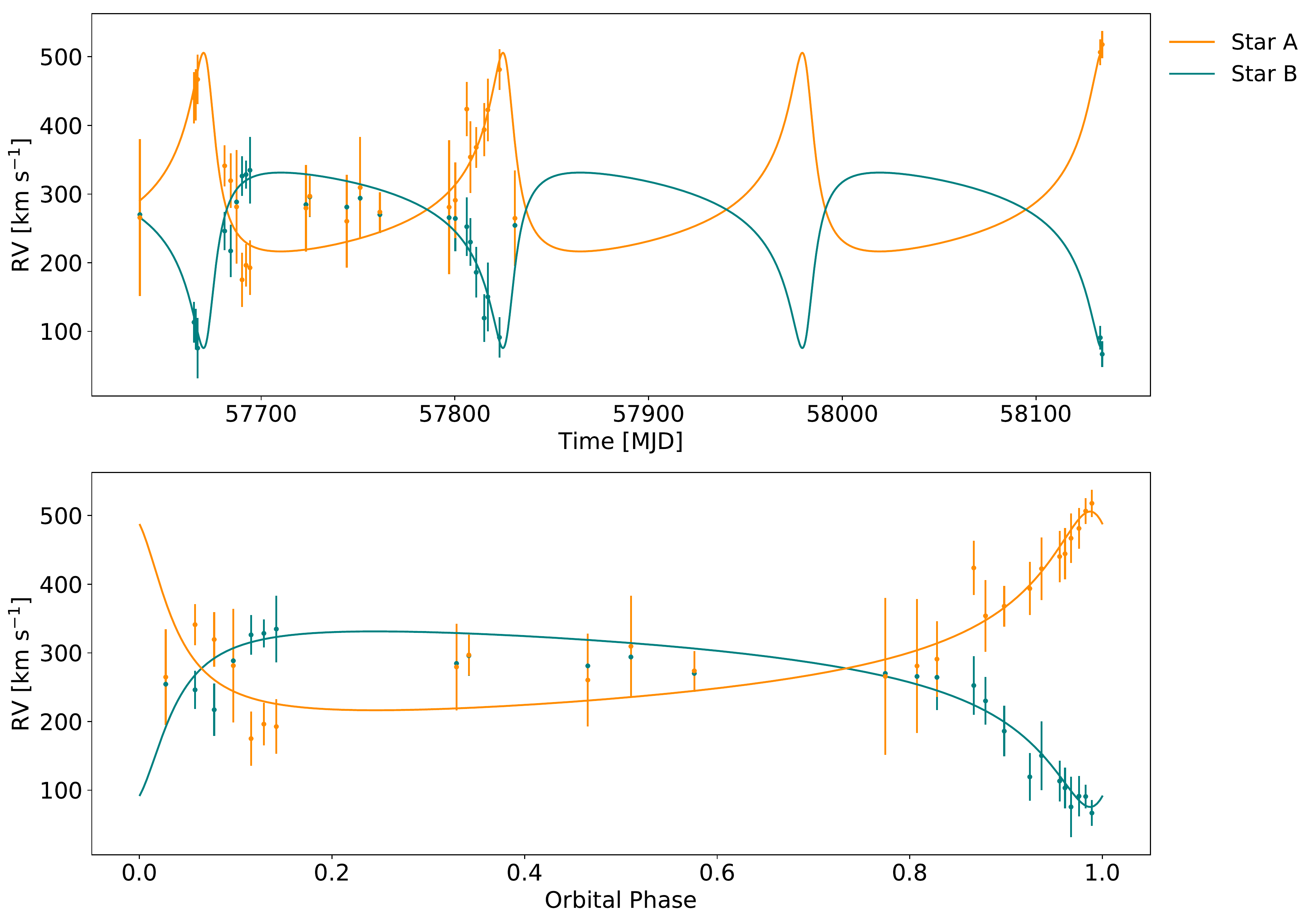}
	\caption{Best fitting radial-velocity curve for the VLT/UVES data providing the parameters for solution U1.}
	\label{fig:rvfit_uves}
\end{figure*}

\section{Orbital Properties -- Solution U2}
\label{appen:solU2}

\begin{figure*}
	\includegraphics[width=\textwidth]{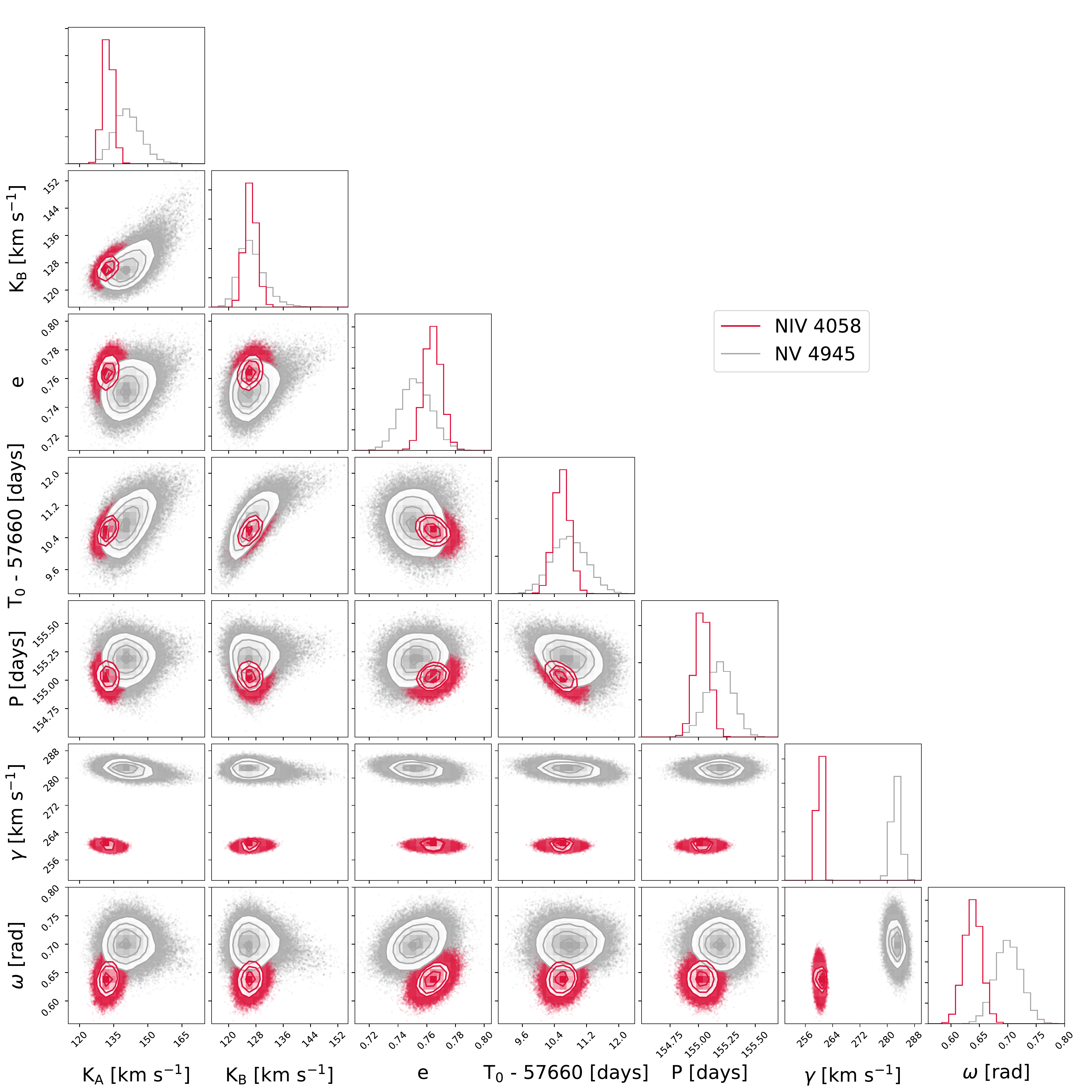}
	\caption{Corner plot showing a comparison of the posterior probabilities for the automated template fitting of the N\,{\sc iv} 4058 emission line (pink contours) and N\,{\sc v} 4945 emission line (grey contours) using the VFTS 682 template from \citet{bes2014}, where $K_{\text{A}}$ and $K_{\text{B}}$ are the semi-amplitudes of the velocities for star A and star B respectively, e is the eccentricity, T$_{0}$ is the time of periastron, P is the orbital period, $\gamma$ is the systemic velocity and $\omega$ is the longitude of the periastron. The results of this fitting were combined to produce solution U2}
	\label{fig:cornerplot_auto}
\end{figure*}


\bsp	
\label{lastpage}
\end{document}